\documentclass[sigconf]{acmart}

\usepackage[T1]{fontenc}
\usepackage{aecompl}
\usepackage{booktabs}
\usepackage{amsmath,amsfonts}
\newtheorem{definition}{Definition}

\usepackage[ruled,linesnumbered]{algorithm2e}
\usepackage{algpseudocode}

\SetKwFor{For}{for}{do}{} 
\SetKwIF{If}{ElseIf}{Else}{if}{then}{else if}{else}{} 
\SetKwFor{ForEach}{for each}{do}{} 
\SetKwFor{While}{while}{do}{} 

\algtext*{End} 

\usepackage{graphicx}
\usepackage{subfigure}
\usepackage{textcomp}
\usepackage{framed}
\usepackage{bm}
\usepackage{enumerate}
\usepackage{tcolorbox}
\usepackage{xcolor}
\usepackage{arydshln}
\usepackage{multirow}
\usepackage{hyperref}
\usepackage[export]{adjustbox}
\usepackage{enumitem}
\usepackage{microtype}
\usepackage{balance}
\usepackage{fancyhdr}
\usepackage{makecell}
\usepackage{todonotes}

\setitemize[1]{itemsep=0pt,partopsep=0pt,parsep=\parskip,topsep=5pt}

\def\BibTeX{{\rm B\kern-.05em{\sc i\kern-.025em b}\kern-.08em
    T\kern-.1667em\lower.7ex\hbox{E}\kern-.125emX}}

\newcommand{\tool}[1]{\textsc{#1}\xspace}
\newcommand{\sysname}{\tool{FairAgent}}

\AtBeginDocument{%
  \providecommand\BibTeX{{%
    \normalfont B\kern-0.5em{\scshape i\kern-0.25em b}\kern-0.8em\TeX}}}

\copyrightyear{2025}
\acmYear{2025}
\setcopyright{acmlicensed}
\acmConference[SIGIR '25]{Proceedings of the 48th
International ACM SIGIR Conference on Research and Development in
Information Retrieval}{July 13--18, 2025}{Padua, Italy}
\acmBooktitle{Proceedings of the 48th International ACM SIGIR Conference on
Research and Development in Information Retrieval (SIGIR '25), July 13--18,
2025, Padua, Italy}
\acmDOI{10.1145/3726302.3729969}
\acmISBN{979-8-4007-1592-1/2025/07}
\begin{document}
\title{Enhancing New-item Fairness in Dynamic Recommender Systems}

\author{Huizhong Guo}
\orcid{0009-0004-0011-8612}
\affiliation{%
  \institution{Zhejiang University}
  \city{Hangzhou}
  \country{China}
}
\email{huiz_g@zju.edu.cn}

\author{Zhu Sun}
\orcid{0000-0002-3350-7022}
\affiliation{%
  \institution{Singapore University of Technology and Design}
  \country{Singapore}
}
\email{sunzhuntu@gmail.com}

\author{Dongxia Wang}
\orcid{0000-0001-9812-3911}
\authornote{Corresponding authors.}
\affiliation{%
  \institution{Zhejiang University}
  \institution{Huzhou Institute of Industrial Control Technology}
  \city{Hangzhou}
  \country{China}
}
\email{dxwang@zju.edu.cn}

\author{Tianjun Wei}
\orcid{0000-0001-7311-7101}
\authornotemark[1]
\affiliation{%
  \institution{Nanyang Technological University}
  \country{Singapore}
}
\email{tjwei2-c@my.cityu.edu.hk}

\author{Jinfeng Li}
\orcid{0000-0002-6400-740X}
\affiliation{%
  \institution{Alibaba Group}
  \city{Hangzhou}
  \country{China}
}
\email{jinfengli.ljf@alibaba-inc.com}

\author{Jie Zhang}
\orcid{0000-0001-8996-7581}
\affiliation{%
  \institution{Nanyang Technological University}
  \country{Singapore}
}
\email{zhangj@ntu.edu.sg}
\begin{abstract}
New-items play a crucial role in recommender systems (RSs) for delivering fresh and engaging user experiences.
However, traditional methods struggle to effectively recommend new-items due to their short exposure time and limited interaction records, especially in \textit{dynamic recommender systems} (DRSs) where \textbf{new-items get continuously introduced} and \textbf{users' preferences evolve over time}.
This leads to significant unfairness towards new-items, which could accumulate over the \textbf{successive model updates}, ultimately compromising the stability of the entire system.
Therefore, we propose \sysname, a reinforcement learning (RL)-based new-item fairness enhancement framework specifically designed for DRSs.
It leverages knowledge distillation to extract collaborative signals from traditional models, retaining strong recommendation capabilities for old-items.
In addition, \sysname introduces a novel reward mechanism for recommendation tailored to the characteristics of DRSs, which consists of three components: 1) a \textbf{new-item exploration reward} to promote the exposure of dynamically introduced new-items, 2) a \textbf{fairness reward} to adapt to users' personalized fairness requirements for new-items, and 3) an \textbf{accuracy reward} which leverages users' dynamic feedback to enhance recommendation accuracy.
Extensive experiments on three public datasets and backbone models demonstrate the superior performance of \sysname. 
The results present that \sysname can effectively boost new-item exposure, achieve personalized new-item fairness, while maintaining high recommendation accuracy.
\end{abstract}



\maketitle

\section{Introduction} 
Recommender systems (RSs) learn user preferences based on historical behavior, continuously providing high-quality items to users~\cite{du2024disentangled, wei2024fpsr+}. 
These systems are widely used across various domains, such as e-commerce~\cite{wang2024amazon}, job recommendation~\cite{du2024enhancing, du2025quasi}, and short-video services~\cite{ge2024short}, providing immense convenience to users and driving significant benefits for platforms and item providers.

In real-world scenarios, new-items are continuously introduced over time, user preferences for these items evolve dynamically, and RSs need to continuously collect user interactions for updates. 
However, in this dynamic recommender systems (DRSs), new-items suffer from limited exposure time, making it difficult to gather sufficient interaction data.
This results in traditional recommendation models failing to effectively learn representations for them, leading to a bias toward over-recommending old-items~\cite{liu2023mitigating,zhou2023adaptive} and exhibiting unfairness toward new-items. 
Such unfairness is further amplified through the dynamic feedback loops of DRSs~\cite{yang2023rectifying}, ultimately compromising the long-term stability of the entire system.

Existing studies have recognized the importance of addressing the issue of new-item fairness~\cite{zhu2021fairness,guo2024configurable}. 
However, these studies fail to account for the dynamic nature of DRSs, leaving the challenge of addressing new-item fairness in DRSs unresolved.
\emph{1) First, the continuous introduction of new-items in DRSs requires optimization objectives to adapt over time.} Existing methods are primarily designed for static scenarios~\cite{zhu2021fairness, guo2024configurable}, meaning they lack the flexibility to adjust to dynamic changes in item pools and user interactions. This inability to accommodate evolving conditions limits their effectiveness in achieving sustained improvements in new-item fairness.
\emph{2) Second,  users' personalized preferences for new-items evolve over time.}
Existing studies on item fairness neglect users' personalized preferences for new-items and also fail to account for the dynamic evolution of these preferences in DRSs.
\emph{3) Third, users provide ongoing interaction feedback that directly influences subsequent model updates.}
To maintain recommendation accuracy, it is essential to dynamically incorporate this feedback into model.
Although some studies have addressed fairness issues in DRSs~\cite{morik2020controlling}, such as mitigating popularity bias~\cite{zhu2021dynamic, wei2023collaborative} and ensuring user-side fairness~\cite{yoo2024ensuring}, they overlook the critical issue of fairness for new-items. 
This aspect is essential for providing fresh user experiences and ensuring the long-term stability of DRSs.
Consequently, the issue of new-item fairness in DRSs remains an open research problem that requires further exploration.

Therefore, we propose \sysname, a reinforcement learning (RL)-based new-item fairness enhancement framework to tackle new-item fairness challenges in DRSs.
Inspired by the idea of knowledge distillation (KD)~\cite{liu2021top}, \sysname inherit pre-trained embeddings from traditional recommendation models, preserving their strong ability to recommend old-items.
On this basis, \sysname specifically designs novel reward strategies tailored for recommending new-items in DRSs.
To address the continuous introduction of new-items in DRSs, \sysname incorporates a \textbf{\textit{new-item exploration reward}} to consistently promote the exposure of newly introduced items, tackling the challenge of limited interaction data for new-items.
To adapt to the dynamic changes in users' personalized preferences for new-items, \sysname introduces a \textbf{\textit{fairness reward}} that dynamically adjusts strategies to enhance fairness, ensuring recommendations align with users' evolving needs. 
Notably, this reward enables \sysname to cater to personalized fairness requirements across different users, providing a more tailored and user-centric recommendation.
Finally, to effectively leverage ongoing user interaction feedback, \sysname introduces an \textbf{\textit{accuracy reward}} designed to maintain recommendation accuracy by ensuring users are presented with items that align with their changing interests.
This addresses the challenge of balancing fairness with accuracy in DRSs.
By integrating these designs, \sysname effectively addresses the key challenges of DRSs, which significantly increases new-item exposure, enhances new-item fairness by considering users' personalized preferences, and maintains high-accuracy recommendations for both new and old items.
Moreover, \sysname can integrate with any existing recommendation models to enhance new-item fairness, offering significant practical value.

In summary, our work makes the following contributions:
\begin{itemize}[leftmargin=0.4cm]
    \item We are the first to introduce the research problem of addressing new-item fairness in DRSs, emphasizing the the critical role of new-items in maintaining the stability of DRSs.
    \item We propose a dynamic RL-based new-item fairness enhancement framework that addresses three key challenges in DRSs: the continuous introduction of new-items, the dynamic evolution of user preferences, and the need for regular model updates. By tackling these challenges, \sysname effectively mitigates unfairness accumulation within feedback loops of DRSs.
    \item We conducted extensive experiments on three public datasets and backbone models. The results demonstrate that \sysname can effectively increase new-item exposure, enhance new-item fairness while maintaining high recommendation accuracy.
    \item We have released the code\footnote{\url{https://github.com/Grey-z/FairAgent}}, establishing \sysname as an open-source tool that can be integrated into any existing recommendation models.
\end{itemize}

\section{Related work}
\textbf{New-item Fairness.}
Ensuring fairness for new-items without prior interaction is crucial in DRSs to enable equal opportunities for all item providers~\cite{zhu2021fairness, guo2024configurable}.
The work~\cite{zhu2021fairness} examines fairness in cold-start scenarios, formalizing it with equal opportunity and Rawlsian Max-Min fairness. It proposes a post-processing framework with two models to enhance fairness among new-items but overlooks unfairness between new and old items. This work~\cite{guo2024configurable} introduces a new-item exposure fairness definition considering item entry-time and presents a framework to address new-item fairness in RSs. Another approach tackles new-item fairness by addressing unfairness caused by varying interaction counts across items, with new-items being least interacted with. For instance, the inverse propensity scoring method~\cite{schnabel2016recommendations} adjusted the training loss by re-weighting interactions according to the inverse of item popularity. Causal intervention methods~\cite{zhang2021causal,wang2021deconfounded} aimed to mitigate the negative effects of popularity on prediction scores, while regularization-based techniques~\cite{zhu2021popularity,rhee2022countering} incorporated fairness constraints into the training loss to balance predictive scores across items. 
However, all of these work overlook the accumulation of new-item unfairness in dynamic feedback loops of DRSs.

\noindent
\textbf{Cold-start Recommender Systems.} 
The cold-start recommendation problem aims to improve a system’s ability to deliver relevant recommendations for new users or items. 
For new-item cold-start scenarios, existing research primarily follows two technical paradigms. 
The first leverages auxiliary item contents, such as category labels, textual descriptions, to reduce reliance on ID embeddings and instead utilize richer semantic features~\cite{volkovs2017dropoutnet, chen2022generative,Huang2023aligning}. 
The second exploits graph-based structures~\cite{du2022metakg, kim2024content, wang2024warming}, including user–item interaction graphs and knowledge graphs, to uncover high-order relational patterns that enhance recommendation quality for new-items.
In this work, we also address the challenge of recommending new-items, with a particular focus on a novel and practical dimension—ensuring fairness in the exposure competition between new and existing items when user attention is limited. 
To support our investigation, we employ state-of-the-art cold-start method~\cite{Huang2023aligning} as the backbone model.

\noindent
\textbf{RL-based Recommender Systems.}
Reinforcement learning has been extensively studied and applied in RSs, offering a powerful framework to optimize long-term user engagement and system objectives~\cite{chen2024opportunities, xin2020self}.
The work~\cite{zhao2018recommendations} proposes a RL-based RS that optimizes strategies through continuous user interaction, effectively incorporating both positive and negative feedback. 
The work~\cite{zheng2018drn} introduces a Deep Q-Learning framework for news recommendation, modeling future rewards to address dynamic user preferences and news features.
And the work~\cite{liu2021top} proposes a top-aware recommender distillation framework that uses RL to refine recommendation rankings, prioritizing top positions to improve user engagement.
Inspired by the work~\cite{liu2021top}, we also leverage KD to inherit traditional model's well-learned information about users and old-items.
The aforementioned works confirm the effectiveness of applying RL in RSs. In this work, we leverage RL techniques to enhance new-item fairness.

\section{Preliminaries}
\begin{figure}[t]
    \setlength{\abovecaptionskip}{6pt}   
    \centering
    \includegraphics[width=0.45\textwidth]{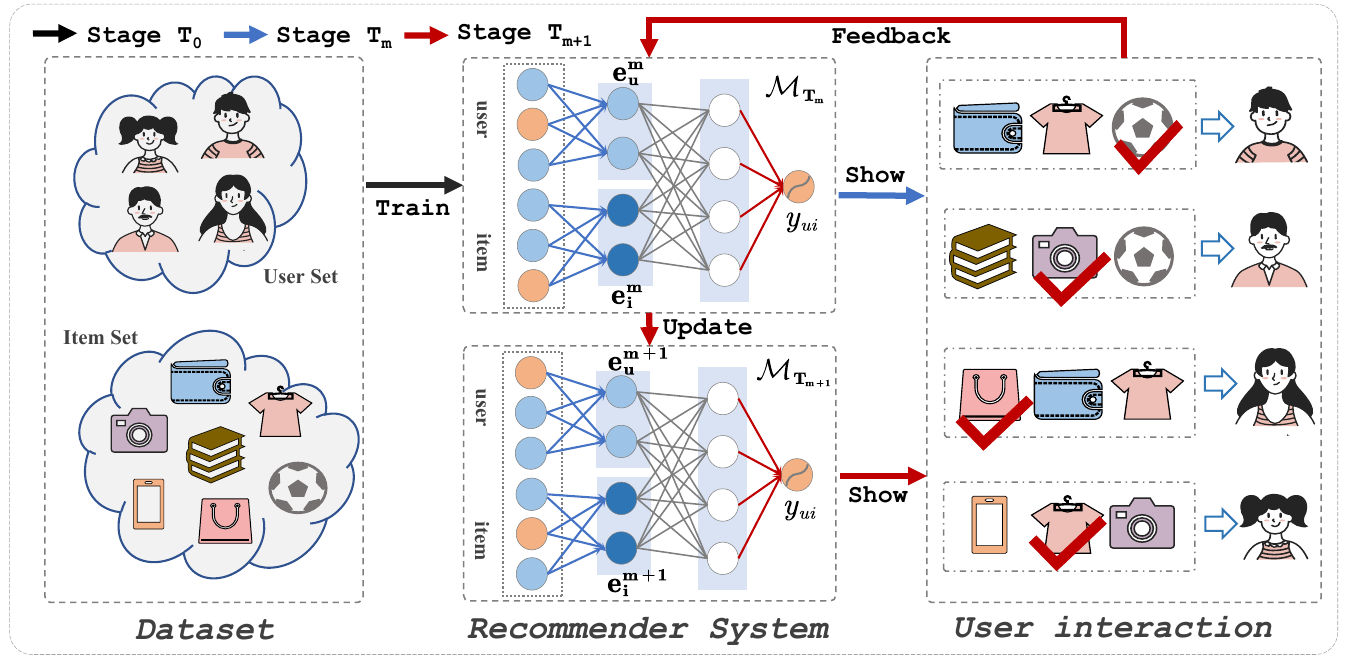}
            \caption{Dynamic recommender system.}
    \label{fig:dynamic_rs}
    \vspace{-0.45cm}
\end{figure}

As illustrated in Fig~\ref{fig:dynamic_rs}, we begin with a concise overview of the basic process of dynamic recommendation scenarios which involves multiple temporal stages~\cite{morik2020controlling, yoo2024ensuring}.
Suppose the DRS starts from $T_0$, where there are a set of users $\mathcal{U}$ and an item set $\mathcal{V}_{T_0}$. 
We collect the user-item interaction data before $T_0$ from all users, denoted as $\mathcal{D}_{T_0} = \{ (u, v) | u \in \mathcal{U}, v \in \mathcal{V}_{T_0}\}$, to train the initial recommendation model, $\mathcal{M}_{T_0}$. 

In each of the following recommendation stage $T_m$, a set of new-items $\mathcal{V}_{T_m}^n$ is introduced into the DRS\footnote{In this work, we focus on fairness from the perspective of new-items. To avoid potential adverse effects, we temporarily exclude the introduction of new users.}.
The updated item set is expressed as:
\begin{equation*}
\footnotesize
\mathcal{V}_{T_m} = \mathcal{V}_{T_{m-1}} \cup \mathcal{V}_{T_m}^n, m = 1,2,\dots, M.
\end{equation*}
$\mathcal{M}_{T_{m}}$ takes user-item pairs $(u, v)$, where $u \in \mathcal{U}, v \in \mathcal{V}_{T_m}$, as input to predict user preference probability $y_{uv} = \mathbf{e}_{u}^{m} \cdot \mathbf{e}_{v}^{m}$, where $\mathbf{e}_{u}^{m}$ and $\mathbf{e}_{v}^{m}$ are the embedding vectors learned by $\mathcal{M}_{T_{m}}$. 
A higher value of $y_{uv}$ signifies a stronger preference of user $u$ for the item $v$.
We consider the top-$K$ recommendation task, where the $K$ items with the highest preference probability are included in the recommendation list, denoted as $L_{u}$, which is then presented to user $u$.

After receiving recommendations, users would typically interact with certain displayed items based on their preferences, such as clicking, purchasing, or adding them to a wish-list. 
Notably, affected by the position bias, items ranked higher on the list tend to receive more exposure, thereby is more probable to get user interactions.
We define \( y_{uv}^* \) as the true preference of user \( u \) for item \( v \), with \( y_{uv}^*=1 \) indicating he/she is willing to interact with the item after observing it and \( y_{uv}^*=0 \) indicating otherwise.
Formally, we model user interaction behavior as follows~\cite{agarwal2019estimating, morik2020controlling}:
\begin{equation}
\footnotesize
    \hat{y}_{uv} =\left\{
    \begin{array}{cc}
        y_{uv}^* \cdot P_{obe}(r_{uv} | v \in L_{u}) & \text { if } r_{uv} \leq K \\
        0 & \text { otherwise }
    \end{array}\right. .
\end{equation}
The observe probability $P_{obe}(r_{uv})$ that represents the probability of $u$ observing the item $v$ ranked at position $r_{uv}$, is modeled as:
\begin{equation}
\footnotesize
p(r_{uv}) \sim \text{Bernoulli}\left(\frac{1}{\log_2(r_{uv} + 1)}\right).
\end{equation}
Noted that the observe probability decreases logarithmically with the value of the rank position.

At the start of the next recommendation stage \( T_{m+1} \), user interaction data from the previous stage, $\mathcal{D}_{T_m} = \{ (u, v, \hat{y}_{uv})|u \in \mathcal{U}, v \in \mathcal{V}_{T_m}\}$, is collected to update the model parameters: $\mathcal{M}_{T_{m}} \rightarrow \mathcal{M}_{T_{m+1}}$. 
The updated model will generate the recommendation lists for users at the next recommendation stage.
Notably, only items appearing in the top-\( K \) recommendation list of a user have the opportunity to receive feedback data, which consequently influences the direction of model updates in the subsequent stages.
This highlights the need to ensure item (exposure) fairness in a DRS, as unfairness may accumulate over stages and eventually lead to system instability, as we will analyze in the following.

\section{New-item Fairness Concern in DRSs}
\label{section:fairness_concern}
In this section, we perform a series of data analyses to reveal exposure fairness concerns of new-items in the existing recommendation models, such as collaborative filtering based models like Matrix Factorization (MF)~\cite{rendle2012bpr}, LightGCN~\cite{he2020lightgcn} and cold-start based models like ALDI~\cite{Huang2023aligning} on the Steam dataset~\cite{o2016condensing}. 
We begin by simulating multiple recommendation stages in DRSs. 
We split the dataset into training and testing sets in a 1:1 ratio. 
The training set is then used to train the backbone model, while the testing set is divided into five subsets to be used in the five sequential stages to simulate a DRS. 
Items are grouped into ten sets based on their appearance time. 
Item sets ($\mathcal{V}^{tr}_1-\mathcal{V}^{tr}_5$) are included in the training set, while item sets ($\mathcal{V}^{te}_1-\mathcal{V}^{te}_5$) are introduced into the system at the start of their corresponding testing stage. 
For instance, items in $\mathcal{V}^{te}_2$ only become available when testing stage $2$ begins and remains accessible in the subsequent stages.
We train the backbone model follows the standard procedures outlined in the public library~\cite{sun2020are, sun2022daisyrec}. 
Detailed experimental settings are provided in Section~\ref{sec:experiments}.

\begin{figure}
    \setlength{\abovecaptionskip}{1pt}
    \centering
    \includegraphics[width=0.45\textwidth]{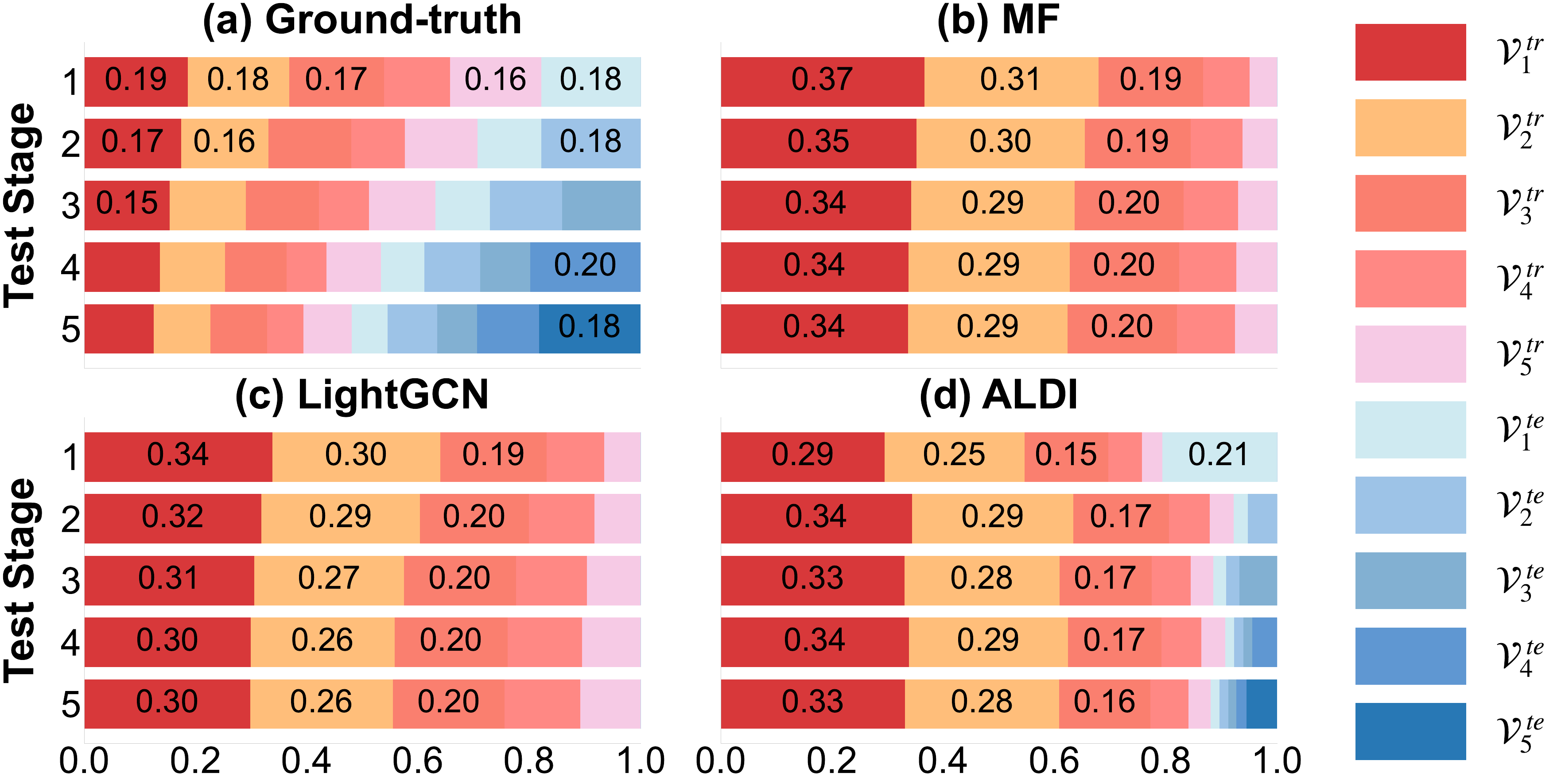}
    \caption{Proportion of items from different sets appearing in the ground-truth and recommendation sets generated by various backbone models.} 
    \label{fig:data-driven}
    \vspace{-0.35cm}
\end{figure}
The findings are presented in Fig~\ref{fig:data-driven}.
In each subfigure, the y-axis shows the results across different recommendation stages, while the x-axis represents different item sets. 
Specifically, red (or orange) shades represent old-items from the training sets, while blue shades indicate new-items from each testing set.
The numbers (or lengths) of the bars reflects the proportion of each item set in the ground truth (Fig~\ref{fig:data-driven}(a)) or in the recommendation sets generated by various backbone models (Fig~\ref{fig:data-driven}(b)-(d)).

From Fig~\ref{fig:data-driven}(a), we observe that new-items introduced in each stage consistently attract a notable share of users in the following stages. 
By test stage 5, they collectively accounted for more market share ( 51\% ) than the old-items.
We could expect that over time new-items may gradually replace more and more old ones in user interaction, which actually complies with lots of applications in reality such as in news/short video/e-commerce platform.
This implies that:
\textbf{\emph{In general, users have increasing interest in exploring new-items over time.}}

Comparing Fig~\ref{fig:data-driven}(b)-(d) to Fig~\ref{fig:data-driven}(a), we observe that in each stage, recommendations of MF, LightGCN and ALDI all deviate from what users actually prefer (i.e. proportion of items distributes differently), and that deviation gets worse along with the stages. 
While under ALDI, new-items can take a bit of market share (though still far from the ground-truth), there is barely chance for them under MF and LightGCN. 
The primary cause of this phenomenon is the lack of user interaction data for new-items. 
Models like MF and LightGCN struggle to capture sufficient information about new-items, resulting in a bias toward overexposing old-items with more interactions. 
While cold-start models like ALDI utilize additional content information to improve the representation of new-items, their exposure remains highly unfair compared to old-items. 
This unfairness is further amplified by the dynamic feedback loops in DRSs, ultimately leaving new-items with minimal exposure.
\textbf{\emph{This highlights a critical issue: under existing RS models, new-items are at a significant disadvantage when competing with old-items for limited exposure, failing to align with actual user needs. It underscores the importance of improving new-item exposure and fairness to enhance the stability and sustainability of dynamic recommender systems.}} 

\section{The Framework of \sysname}
Aiming for improving the exposure fairness of new-items, in this section, we propose a RL-based new-item fairness enhancement framework, \sysname. 
In this section, we first introduce the relevant fairness definitions and evaluation metrics related to fair exposure of new-items.
Next, we elaborate on the design of the proposed \sysname framework, and how it explores new-items and maintains fairness between new and old items.

\begin{figure*}[t]
\setlength{\abovecaptionskip}{2pt}
\centerline{
\includegraphics[width=0.7\textwidth]{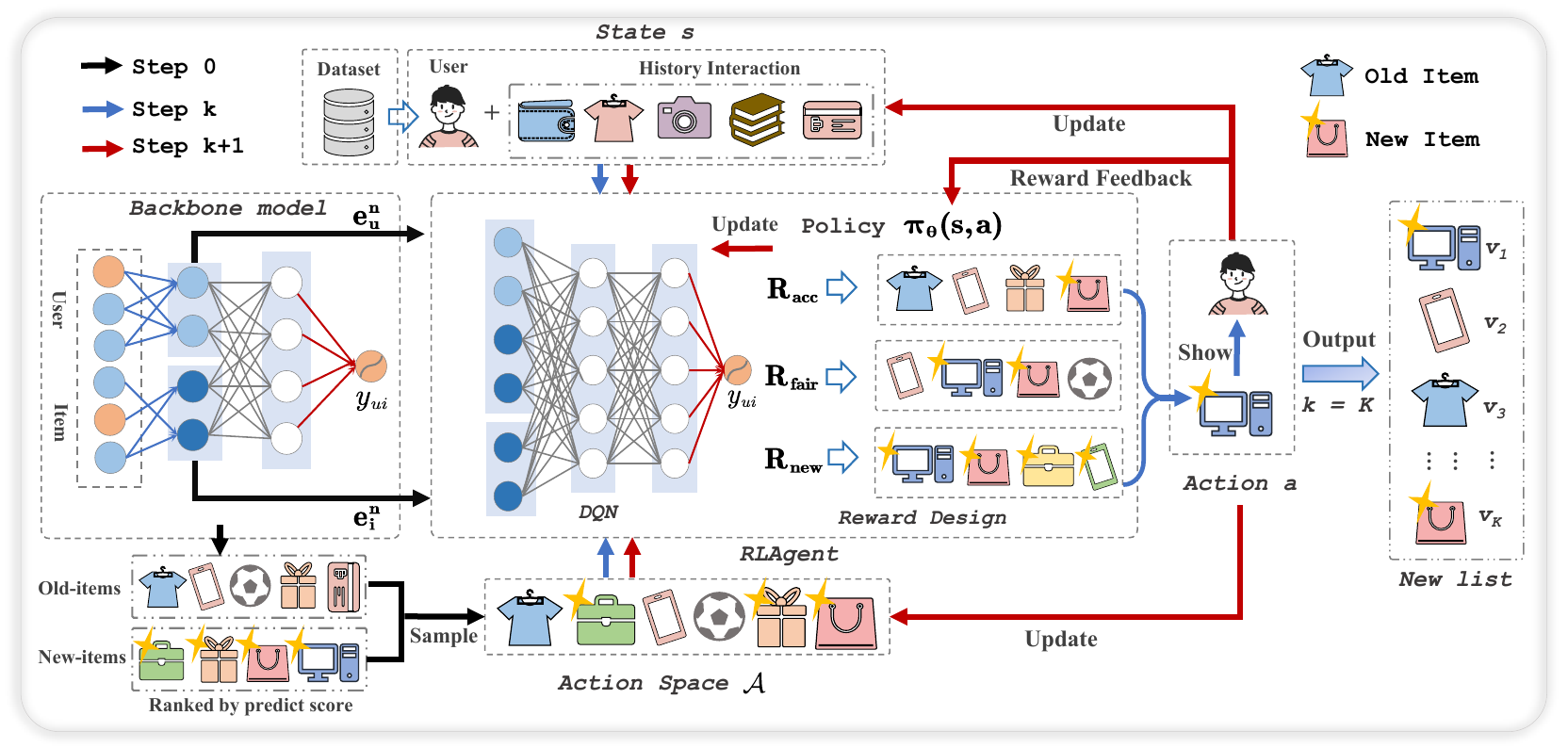}}
\caption{Our proposed RL-based new-item fairness enhancement framework, \sysname.}
\label{fig:framework}
\vspace{-0.25cm}
\end{figure*}
\label{sec:framework}

\subsection{User-Level New-Item Exposure Fairness}
In DRSs, item providers are primarily concerned with whether their items are effectively exposed to users, as this is a prerequisite for potential user interactions.
We define exposure resources that item \( v \) receives in recommendation list $L_u$ during stage \( T_m \) as~\cite{morik2020controlling}:
\begin{equation}
 \label{eq:exp}
 \footnotesize
    Exp_m\left(L_u,v\right)=\frac{\mathbb{I}\left(v \in L_u\right)}{\log _2\left(r_{uv} +1\right)},
\end{equation}
where function $\mathbb{I}(x)$ serves as an indicator, returning $1$ if $x$ is true and $0$ otherwise. 
\( r_{uv} \) denotes the rank position of item \( v \) within \( L_u \). 
Items ranked higher in the list receive more exposure and are more likely to attract user interactions.

We have observed in Section~\ref{section:fairness_concern} that with new-items entering a system over time, proportion of user preference of old and new items changes dynamically. 
As shown in Fig~\ref{fig:data-driven}, typically for new-items, those that entered more recently tend to attract more users, and for old-items, those relatively older ones attract more users.
This enlightens us that to investigate whether exposure resources are fair between old and new items, their entry time should be considered.
There exists an time-based item fairness metric (TGF) proposed in~\cite{guo2024configurable} that takes this into account. It weights items based on their entry time in measuring the exposure disparity between old and new items. 
\begin{equation}
\label{eq:tgf}
\footnotesize
\aligned
    &TGF(L_u) = \frac{1}{|\mathcal{V}^o_{u}|}\sum_{p=1}^{|\mathcal{V}^o_{u}|} w_p^o \cdot Exp_m(L_u,v_p^o) - \frac{1}{|\mathcal{V}^n_{u}|}\sum_{q=1}^{|\mathcal{V}^n_{u}|} w_q^n \cdot Exp_m(L_u,v_q^n),
\endaligned
\end{equation}
\begin{equation}
\label{eq:weight}
\footnotesize
    w_p^o = |\mathcal{V}^o_{u}| + 1 - p, \ \ w_q^n = 1 + (q - 1) \cdot \frac{|\mathcal{V}^o_{u}| - 1}{|\mathcal{V}^n_{u}| - 1},
\end{equation}
where \( \mathcal{V}^o_{u} \) and \( \mathcal{V}^n_{u} \) represent the sets of old and new items, respectively, within \( L_{u} \) at stage \( T_m \). 
Items are ordered in descending sequence based on their entry time into the DRS. 
Older items in \( \mathcal{V}^o_{u} \) with smaller indices \( p \) are assigned larger weights \( w_p^o \) considering their accumulated larger user base and entitlement to higher exposure resources. 
Newer items in \( \mathcal{V}^n_{u} \) with larger indices \( q \) are assigned larger weights \( w_q^n \) to make up for their disadvantages in collecting interactions.

We want to follow the idea of TGF since it aligns with our observations, but there is a problem.
Taking the average over all users, TGF cannot account for the personalized preferences for new and old items, which could vary between different users. 
To address this limitation, we propose a user-level personalized fairness metric:
\begin{definition}User-level personalized New-item Fairness (UNF).

For user $u \in \mathcal{U}$, let \( H_{u} \) denote his/her historical interaction list, while \( L_{u} \) still denotes recommendation list the user obtained. 
UNF is defined as the divergence between the TGF calculated by recommendation results and that of user's historical interactions.
\begin{equation}
\label{eq:dkl}
\footnotesize
   UNF = \frac{1}{|\mathcal{U}|} \sum_{u \in \mathcal{U}} \left[\text{TGF}(L_{u}) - \text{TGF}(H_{u})\right]^2.
\end{equation}
The value of \( \text{TGF}(L_{u}) \) reflects the exposure distribution of new and old items in the recommendation results, while \( \text{TGF}(H_{u}) \) represents the user's historical preference for that, computed based on their past interactions.
\end{definition}
A smaller value of UNF indicates that the recommendation results align more closely with the user's preference for new and old items, making the system \emph{fairer at the individual user level}.

\subsection{Fairness Enhancement with RL Framework} 
In this section we present the design details of our RL-based new-item fairness enhancement framework, \sysname. 
As shown in Fig~\ref{fig:framework}, \sysname initializes from a backbone model, selects the candidate items based on user interaction and a carefully designed reward mechanism, and then produces a fairer top-\( K \) recommendation list.
A core characteristic of \sysname lies in its design of the reward mechanism, which offers three significant advantages, 1) appropriate new-item recommendation rate and 2) fairer exposure of new-items and 3) high recommendation accuracy.
The implementation details will be elaborated in the following sections.

\subsubsection{Setting of RL-based DRS}
Following the paradigm of RL-based work~\cite{zheng2018drn, liu2021top}, we utilize the following settings in \sysname:
\begin{itemize}[leftmargin=0.3cm]
    \item \textbf{State} $\mathbf{s}_{u}^t$ refers to a vector that captures the historical preferences of a specific user \( u \) at step $t$.
    It comprises of the embeddings of the users along with most recent \( N \) items they have interacted with, represented as $\mathbf{s}_{u}^t = [\mathbf{e}_u, \mathbf{e}_{v_1}, \mathbf{e}_{v_2}, \cdots, \mathbf{e}_{v_N}]$.
    \item \textbf{Action} $\mathbf{a}_{u}^t$ refers to the item selected for user \( u \) at step \( t \) from the action space $\mathcal{A}_{u}^t$. 
    \item \textbf{Reward} $r_{u}^t$ represents the benefit obtained by selecting action $\mathbf{a}_{u}^t$ given the state $\mathbf{s}_{u}^t$.
\end{itemize}

\textbf{Deep Q-Network (DQN)} \cite{mnih2013playing} is a RL algorithm that integrates Q-learning with deep neural networks, enabling it to approximate the Q-value function for high-dimensional state-action spaces. 
In DRSs, DQN learns the optimal policy $\pi_{\theta}(\mathbf{s}^t_u,\mathbf{a}^t_u)$ for recommendations by modeling user-item interactions as a Markov Decision Process (MDP). 
The Q-value, \( Q(\mathbf{s}_{u}^t, \mathbf{a}_{u}^t) \), represents the expected cumulative reward of selecting an item \( v  \) (action $\mathbf{a}_{u}^t$) given the user’s  preferences (state \( \mathbf{s}_{u}^t \)), guiding the system to recommend items that maximize user satisfaction or system objectives.

\subsubsection{Detailed Step to Construct \sysname}
Based on the aforementioned settings, we now provide a detailed explanation of the steps involved to construct \sysname, as illustrated in Algorithm~\ref{algo:training_process}.

(1) \textbf{Initialize Model} (lines 1-3): 
\sysname employs DQN, denoted as $\mathcal{Q}_{\theta}$, as its core component to predict user preferences for items.
To accelerate convergence, DQN is initialized with pre-trained embeddings \( \mathbf{e}_{\mathcal{U}} \) and \( \mathbf{e}_{\mathcal{V}_{T_0}} \) from the backbone model $\mathcal{M}_{T_0}$
This enables DQN to retain the backbone model's strong recommendation capability for old-items and build upon it to enhance recommendations for new-items.

(2) \textbf{Construct initial state and action space} (lines 5-11):
We combine the embeddings of the user and the \( N \) items he/she most recently interacted with during the previous recommendation stage to construct the initial state, $\mathbf{s}_{u}^0$.
For the construction of action space $\mathcal{A}_u^0$, we introduced a preference-aware sampling strategy to dynamically adjust the ratio of old and new items.
Specifically, we define a \textit{Bernoulli} distribution over the binary variable "item type" (old or new), parameterized by the ratio of new-items in user's historical interactions \( p_{new} \). Each item added to the action space is sampled from this distribution: with probability \(p_{new}\) it is drawn from the set of new-items, and with probability \( (1-p_{new}) \) it is drawn from the set of old-items. For instance, if the estimated probability of the user preferring a new-item is \( P_{\text{new}} = 0.4 \), the distribution is configured to yield an old-to-new item ratio of 6:4. 
For the selection within the new and old item sets, we adopt the same greedy search strategy: select the item with the highest score remaining in the item set, where the score is calculated from the pre-trained backbone models.
By repeating this process for $K$ steps, we construct a fixed-length initial action space $\mathcal{A}_u^0$ .

(3) \textbf{Choose action} (lines 14-17): 
At each step $t$, \sysname takes an action $\mathbf{a}_{u}^t$ (i.e., selects an item $v$ to user $u$) from action space $\mathcal{A}_u^t$ based on the current user state $\mathbf{s}_{u}^t$. 
During training, we use a strategy combined with $\varepsilon$-greedy exploration. 
Specifically, with a probability of $1 - \varepsilon$, the model selects the item with the highest Q-value from the action space, while with a probability of $\varepsilon$, it randomly selects an item to encourage exploration.
This approach balances exploitation and exploration, allowing the model to avoid local optima and improve its generalization capabilities.

\begin{algorithm}[t]
\footnotesize
\caption{Training Process of \sysname at stage $T_m$}
\label{algo:training_process}
\KwIn{$\mathcal{D}_{T_{m}}$, $\mathcal{M}_{T_{m}}$, $\mathcal{V}_{T_m}$, $\mathcal{U}$, $H_\mathcal{U}$} 
\KwOut{Updated model $\mathcal{Q}_{\theta}, \mathcal{Q}_{\theta^\prime}, L_u$}

\If{$m == 0$}{
    $\mathbf{e}_{\mathcal{U}}, \mathbf{e}_{\mathcal{V}_{T_0}} \leftarrow \mathcal{M}_{T_{0}}$  \tcp*[l]{Get pre-trained embeddings}
    $\mathcal{Q}_{\theta}, \mathcal{Q}_{\theta^\prime} = \text{Initialize}(\mathbf{e}_{\mathcal{U}}, \mathbf{e}_{\mathcal{V}_{T_0}})$ \;
}

\ForEach{$u \in \mathcal{U}$}{
    $\mathbf{s}_{u}^0 = [\mathbf{e}_{u}, \mathbf{e}_{v_1}, \mathbf{e}_{v_2}, \cdots, \mathbf{e}_{v_N}]$ \;
    $P_{new} = \text{GetPreference}(u, H_u)$ \;    \While{$len(\mathcal{A}_u^{0}) < N_{act}$}{
        \If{$\mathbb{I}({x=1| x \sim Bernoulli(P_{new})})$}{
            $\mathcal{A}_u^{0} \leftarrow \text{Sample}(\mathcal{V}_{T_m}^n)$ \tcp*[l]{Sample a new-item.}
        }
        \Else{
            $\mathcal{A}_u^{0} \leftarrow \text{Sample}(\mathcal{V}_{T_{m-1}})$ \tcp*[l]{Sample an old-item.}  
        }
    }
    $L_{u}^0 \leftarrow \varnothing$ \;
    \For{$t \gets 1$ \textbf{to} $K$}{
        \If{\texttt{train} == \texttt{True} \textbf{ and } $ \mathbb{I}({x=1| x \sim Bernoulli(\varepsilon)})$}{
            $\mathbf{a}_{u}^t = \text{Random}(\mathcal{A}_u^t)$ \tcp*[l]{Choose action randomly}
        }
        \Else{
            $\mathbf{a}_{u}^t = \text{ChooseAction}(\mathbf{s}_{u}^t, \mathcal{A}_u^t,\mathcal{Q}_{\theta})$ \;
        }
        $L_u^{t} \leftarrow L_u^{t-1} \ \cup \ \mathbf{a}_{u}^t (v) $ \;
        $r_u^t = \text{GetReward}(\mathbf{s}_{u}^t, \mathbf{a}_{u}^t, \mathcal{V}_{T_m}^n)$ \;
        $\mathcal{A}_u^t = \text{Update}(\mathcal{V}_{T_m})$ \;
        
        \If{$\mathbf{a}_{u}^t \ \texttt{returns positive feedback}$}{
            $\mathbf{s}_u^{t+1} = \text{UpdateState}(\mathbf{a}_{u}^t, \mathbf{s}_u^{t})$ \tcp*[l]{Update state} 
        }
        \Else{
            $\mathbf{s}_u^{t+1} = \mathbf{s}_u^{t}$ \;
        }
        
        $\mathcal{D}_{\text{buffer}} \leftarrow (\mathbf{s}_u^{t}, \mathbf{a}_u^{t}, r_u^{t}, \mathbf{s}_u^{t+1})$ \tcp*[l]{Memory mechanism}
        
        \If{$\text{len}(\mathcal{D}_{\text{buffer}}) > N_{\text{mem}}$}{
            $\mathcal{Q}_{\theta^{\prime}} = \text{Train}(\mathcal{D}_{\text{buffer}})$ \tcp*[l]{Update parameters}
        }
        $\mathcal{Q}_{\theta^\prime} = \text{UpdateNetwork}( 
            \mathcal{Q}_{\theta}$) \;
            \tcp{Update target network every five iterations}
    }
    $L_u \leftarrow L_u^{K}$\;
}
\Return{$\mathcal{Q}_{\theta}, \mathcal{Q}_{\theta^\prime}, L_u$}\;
\end{algorithm}

(4) \textbf{Calculate reward} (lines 18-19):
After taking an action, the selected item \( v \) is added to the recommendation list \( L_u \).
The reward is subsequently calculated based on the updated \( L_u^t \) at step $t$.

(5) \textbf{Update action space and state} (lines 20-24):
After step $t$ completed, both the action space $\mathcal{A}^t_u$ and the state $\mathbf{s}^t_u$ need to be updated for the next step $t+1$. 
For the action space, the selected item $v$ is removed from $\mathcal{A}^t_u$, and another item is sampled and added into $\mathcal{A}^{t+1}_u$, ensuring the action space size remains constant. 
If the selected item $v$ receives user's positive feedback, i.e., $\hat{y}_{uv} = 1$, the state is updated in a similar manner. 
The earliest interacted item in the $\mathbf{s}^t_u$ is removed, and $\mathbf{e}_{v}$ is added to the $\mathbf{s}^{t+1}_u$. If no positive feedback is received, the state $\mathbf{s}^{t+1}_u$ remains the same as $\mathbf{s}^{t}_u$.

(6) \textbf{Update DQN} (lines 25-28)
Inspired by the existing works~\cite{schaul2015prioritized, o2010play, liu2021top}, we utilize two key techniques to improve stability and performance of DQN, including:
1) \emph{Experience Replay}. A buffer is used to store past experiences $(\mathbf{s}_u^{t}, \mathbf{a}_u^{t}, r_u^{t}, \mathbf{s}_u^{t+1})$, which are sampled randomly during training to break the correlation between consecutive updates.
2) \emph{Target Network}. A separate target network $\mathcal{Q}_{\theta^{\prime}}$ is maintained and periodically updated to stabilize the Q-value estimates, preventing rapid oscillations during training.
In \sysname, DQN is adapted to model user decision-making processes and optimize the ranking of items based on a reward mechanism, ensuring accurate and fair recommendations.
Specifically, the parameters $\theta$ of DQN are trained using the following loss function:
\begin{equation}           
\footnotesize
    \mathcal{L}_{\theta}=\mathbb{E}_{\mathbf{s}^t_u, \mathbf{a}^t_u, r^t_u, \mathbf{s}^{t+1}_u}\left[\left(Q_{\text{target}}-Q\left(\mathbf{s}^t_u, \mathbf{a}^t_u ; \theta\right)\right)^2\right],
\end{equation}
where $Q(\mathbf{s}_u^{t}, \mathbf{s}_u^{t})$ represents the Q-value for the current state-action pair, $Q_{\text{target}}$ denotes the target Q-value. For each training step, the target Q-value is defined as:
\begin{equation}
\footnotesize
   Q_{\text{target}} = \mathbb{E}_{\mathbf{s}^{t+1}_u} \left[ r^t_u + \lambda \max_{\mathbf{a}^{t+1}_u} Q(\mathbf{s}^{t+1}_u, \mathbf{a}^{t+1}_u; \theta') \mid \mathbf{s}^t_u, \mathbf{a}^t_u \right],
\end{equation}
where $\theta'$ indicates the parameters of the target network, which are periodically updated from the main Q-network, such as in every 5 iterations. 
The constant $\lambda$, ranging between 0 and 1, determines the balance between current and future rewards.
\sysname updates the recommendation policy $\pi_{\theta}(\mathbf{s}^t_u,\mathbf{a}^t_u)$ to find out the optimal policy parameter $\theta$ that can maximize the expected cumulative rewards.

\subsubsection{Reward Mechanism}
\label{subsubsec:reward}
As previously described, \sysname employs a reward mechanism to generate the recommendation list for all users.
The objective is to produce a refined list that satisfies the following three key properties: 
1) \textit{Appropriate new-item recommendation rate}. 
The continuously introduced new-items receive sufficient exposure to meet users' needs.
2) \textit{Fair exposure allocation for new and old items}. 
Exposure resources are distributed fairly between new and old items, aligning with users' preferences for new-items.
3) \textit{High accuracy}. 
The recommended items closely reflect users' true preferences for both new and old items.

The three rewards below are designed to guide model update to achieve the above properties respectively.

\textbf{New-item exploration reward ($R_{new}$)} aims to encourage the exploration of continuously introduced new-items, thus improving the exposure rate of new-items and ensuring sustainability of the whole DRS.
\begin{equation}
\footnotesize
    R_{new} = \gamma \cdot \mathbb{I}(v \in \mathcal{V}^n_{T_m})+ (1-\gamma) \cdot \mathbb{I}(v \in \mathcal{V}^n_{T_m}) \cdot \mathbb{I}(\hat{y}_{uv}=1)
\end{equation}
We use parameter \( \gamma \) to control the distribution of reward assigned to recommending new-items that have received user positive feedback and that have not.

\textbf{Fairness reward ($R_{fair}$)} aims to adjust the distribution of new and old items in the recommendation list to align with users' evolving preferences for them, thereby achieving personalized new-item fairness.
Let 
\begin{equation}
\footnotesize
\begin{aligned}
UNF_u^t &= \left| TGF(L_{u}^t) - TGF(H_u) \right|, \\
R_{fair} &= \frac{2 \cdot \text{tanh}(UNF_u^t - UNF_u^{t+1})}{1 + \text{tanh}(2)},
\end{aligned}
\end{equation}
where \( L_{u}^{t} \) denotes the generated recommendation list at step \( t \), while \( H_u \) represents the user's historical interaction list, which reflects their historical preferences for new-items. 
The fairness reward \( R_{fair} \) ranges between \((-1, 1)\), taking a positive value when \( TGF(L_{u}) \) aligns with \( TGF(H_u) \) and a negative value otherwise. 
This reward is designed to optimize the recommendation list to better align with the user's true preferences for new-items.

\textbf{Accuracy reward ($R_{acc}$)} leverages ongoing user interaction feedback, dynamically adjusting to ensure the system adapts to evolving preferences and effectively utilizes this feedback to enhance recommendation accuracy over time.
\begin{equation}
\footnotesize
    R_{acc} = \frac{\mathbb{I}(\hat{y}_{uv}=1)}{\text{log}_2(r_{uv} + 1)}
\end{equation}
where $\hat{y}_{uv}=1$ denotes a positive user feedback to item $v$.

Taking accuracy as the basic goal, new-item fairness and exploration as the additional ones, we define the total reward as:
\begin{equation}
\footnotesize
R_{total} = R_{acc} + \alpha R_{fair} + \beta R_{new},
\label{eq:total_reward}
\end{equation}
where \( \alpha \) and \( \beta \) regulate the influence of $R_{fair}$ and $R_{new}$ respectively. 

\section{Experiments}
\label{sec:experiments}
In this section, we conduct extensive experiments on three publicly available datasets and backbone models. 
By addressing the following three research questions consecutively, we demonstrate the superior performance of \sysname in improving exposure rate of new-items, achieving new-item fairness while aligning with user's personalized preference for new-items and maintaining high recommendation accuracy.

\noindent
\noindent
\textbf{RQ1:} Compared to state-of-the-art (SOTA) baselines, can \sysname more effectively improve exposure rate of new-items, address unfairness between new and old items, while maintaining high recommendation accuracy?

\noindent
\textbf{RQ2:} Can \sysname keep up with user's personalized dynamic preferences for old and new items in DRSs?

\noindent
\textbf{RQ3:} How effective are the different reward components in the design of \sysname?

\textbf{Datasets.}
To be align with real-world situations, we selected three publicly available datasets to construct varying DRS scenarios.
Each dataset reflects distinct user behavior patterns.
\textbf{KuaiRec-Small}~\cite{gao2022kuairec} is a dense dataset collected from the Kuaishou platform, containing user interactions with short videos.
In each stage, numerous new-items enter the DRS, and user interest shift towards these items rapidly.
\textbf{KuaiRec-Large}~\cite{gao2022kuairec} is a larger version of the previous dataset. 
This dataset features a DRS with relatively smoother expansion of item set and slower shifts of user interests.
For these two datasets, we filter interactions where users' watch ratio is greater than $1.0$ as positive samples.
\textbf{Steam}~\cite{o2016condensing} is a game dataset containing user interaction data from the Steam gaming platform. 
It encompasses a larger pool of users and items, with relatively stable item set expansion and more consistent changes in user interests.
To create a DRS setting, we split all interaction data into training and test sets at a 1:1 ratio in chronological order. 
We train the backbone model on the entire training set, and utilize the last 20\% to train \sysname and all the baselines. 
The test set is then divided into 5 sets to construct different test stages, each introducing a set of new-items to reflect dynamic introduction of new-items.
For all datasets, we filter out users with fewer than $10$ interactions.

\begin{figure}[t]
\setlength{\abovecaptionskip}{6pt}
\centerline{
\includegraphics[width=0.4\textwidth]{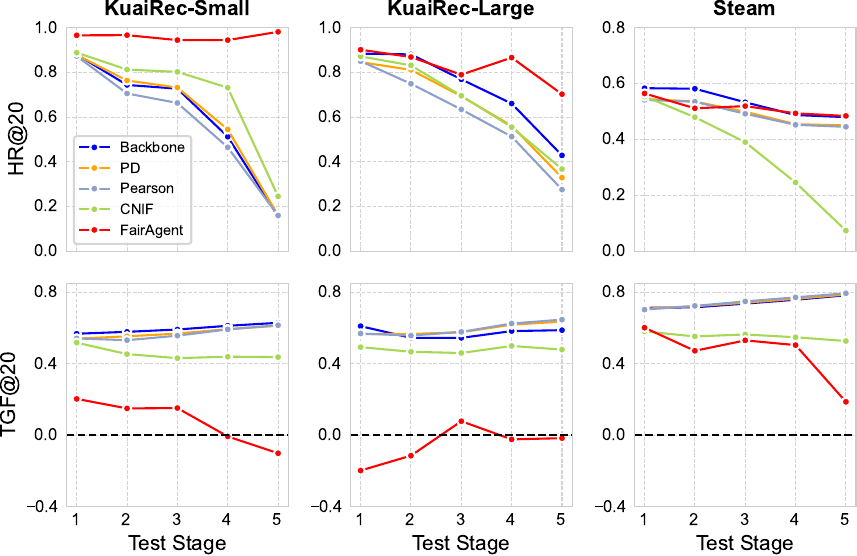}}
\caption{Dynamic changes in HR and TGF metrics across 5 test stages for different methods using the MF backbone.}
\label{fig:result_rq1_mf}
\vspace{-0.4cm}
\end{figure}

\textbf{Backbone Model.}
Following the settings of the existing works~\cite{rhee2022countering,guo2024configurable}, we employ widely used models, including Matrix Factorization (MF)~\cite{rendle2012bpr}, the graph-based recommendation algorithm LightGCN~\cite{he2020lightgcn}, and the cold-start recommendation algorithm ALDI~\cite{Huang2023aligning}, as backbone models to validate the effectiveness of \sysname. 
Noted that both MF and LightGCN rely solely on user-item interaction data, while ALDI requires content information to recommend new-items. 
For fair comparison, we utilize additional content information only when ALDI was used as the backbone model for all the baseline methods. 
We constructed the content information using a pre-trained language model~\cite{reimers-2019-sentence-bert}, extracting features from the titles and categories of short videos (for KuaiRec-Small and KuaiRec-Large), as well as the names and labels of games (for Steam). 

\textbf{Comparison Baselines.}
To the best of our knowledge, this is the first work to address new-item fairness in DRSs. 
We opted to adapt methods designed to enhance item fairness in static scenarios as baselines for comparison. 
\textbf{PD}~\cite{zhang2021causal} employs causal intervention to adjust the final prediction scores based on the amount of each item's interactions, thereby balancing the overall recommendation rate.
\textbf{Pearson}~\cite{zhu2021popularity} introduces a regularization term that incorporates the correlation between prediction scores and item popularity to enhance item fairness during training.
These baselines primarily focus on increasing the recommendation exposure of items with limited interactions, where here new-items are considered as those with no interaction.
\textbf{CNIF}~\cite{guo2024configurable} is the first work that explicitly considers the time at which an item enters a system, and introduces a time-based fairness loss function to optimize new-item fairness during training.
The parameters of these baselines were tuned based on the configurations in the existing work~\cite{guo2024configurable}.

\textbf{Implementation Details.}
We follow the training protocol from the Open-Source Library~\cite{sun2022daisyrec} to train all backbone models with BPR loss. 
Then we employ Bayesian optimization to tune the relevant hyper-parameters, including learning rate, regularization coefficient, and embedding dimension. 
For fair comparison, all experiments use a batch size of $8,192$ and a negative sampling rate of $4$ for training. 
For the parameters mentioned in Section~\ref{subsubsec:reward}, we tune $\alpha$ within $[0.5,2.5]$, $\beta$ within $[0, 1]$, and set $\gamma = 0.1$.

\textbf{Evaluation and Metrics}.
Following the setting of \cite{liu2021top, sun2022daisyrec}, at each test stage, we sample negative items for users to construct a fixed-length candidate set of 1000 items for evaluation.
To measure recommendation accuracy, we adopt Hit Rate (HR)~\cite{sun2022daisyrec,xin2020self} and NDCG~\cite{jarvelin2002cumulated, du2024enhancing}, with larger values indicating better performance. 
We also report TGF (introduced in~\cite{guo2024configurable}) at all stages, to evaluate overall new-item fairness (considering all the users), with larger values denoting more unfair exposure distribution against new-items. 
Furthermore, we use New-item Coverage (NC) to
represent the ratio of new-items in recommendation lists for all users and a trade-off metric $\delta T$ to quantify the balance between fairness improvement and any potential loss in recommendation accuracy following the work~\cite{guo2024configurable}.
A higher $\delta T$ indicates more fairness enhancement and less accuracy loss.
Due to page limitations, we report results for all the metrics with $K=20$. Note that similar conclusions can be drawn for other $K$ values.

\subsection{Experimental Results and Analysis}
\subsubsection{Results of RQ1}

\begin{table*}[tp] 
\setlength{\abovecaptionskip}{6pt}
\setlength\tabcolsep{4pt}
\centering
\caption{Average results across all test stages for different methods on three backbone models.}
\label{tab:average_results}
\resizebox{0.8\textwidth}{!}{
\begin{tabular}{ccccccccccccccccccc}
    \toprule
    \textbf{Method} & \multicolumn{6}{c}{\textbf{MF}}               & \multicolumn{6}{c}{\textbf{LightGCN}}         & \multicolumn{6}{c}{\textbf{ALDI}} \\
\cmidrule(lr){2-7}    
\cmidrule(lr){8-13}
\cmidrule(lr){14-19}
\textbf{K=20} & HR$\uparrow$    & NDCG$\uparrow$  & TGF$\downarrow$   & UNF$\downarrow$   & NC$\uparrow$    & $\%\delta T \uparrow$ & HR$\uparrow$    & NDCG$\uparrow$  & TGF$\downarrow$   & UNF$\downarrow$   & NC$\uparrow$    & \multicolumn{1}{c}{$\%\delta T\uparrow$} & HR$\uparrow$    & NDCG$\uparrow$  & TGF$\downarrow$   & UNF$\downarrow$   & NC$\uparrow$    & $\%\delta T\uparrow$ \\
    \midrule
    \midrule
          & \multicolumn{18}{c}{KuaiRec-Small} \\
\cmidrule{2-19}    \textbf{Backbone} & 0.6035  & 0.2741  & 0.5949  & 0.1252  & 0.0000  &   -  & \underline{0.6379}  & \underline{0.2972}  & \underline{0.5694}  & \underline{0.1156}  & 0.0000  &  -   & 0.8593  & 0.4269  & 0.2448  & 0.0385  & \underline{0.0183}  & -\\
    \textbf{PD} & 0.6154  & 0.2781  & 0.5729  & 0.1195  & 0.0000  & 102.9\% & 0.4749  & 0.2142  & 0.6717  & 0.1425  & 0.0000  & 80.8\% & 0.8163  & 0.4040  & 0.3711  & 0.0691  & 0.0000  & 72.9\% \\
    \textbf{Pearson} & 0.5729  & 0.2584  & 0.5667  & 0.1170  & 0.0000  & 99.8\% & 0.4736  & 0.2132  & 0.6741  & 0.1432  & 0.0000  & 80.5\% & 0.8845  & \underline{0.4558}  & 0.2564  & 0.0420  & 0.0036  & 99.1\% \\
    \textbf{CNIF} & \underline{0.6957}  & \underline{0.3211}  & \underline{0.4552}  & \underline{0.0903}  & 0.0000  & \underline{120.9\%} & 0.4770  & 0.2302  & 0.6458  & 0.1419  & 0.0000  & \underline{82.9\%} & \underline{0.8883}  & 0.4538  & \underline{0.2286}  & \underline{0.0357}  & 0.0054  & \underline{105.1\%} \\
    \textbf{\sysname} & \textbf{0.9600}  & \textbf{0.5082}  & \textbf{0.0793}  & \textbf{0.0061}  & \textbf{0.3003}  & \textbf{197.5\%} & \textbf{0.9413}  & \textbf{0.5025}  & \textbf{0.0866}  & \textbf{0.0124}  & \textbf{0.3306}  & \textbf{182.7\%} & \textbf{0.9591}  & \textbf{0.5540}  & \textbf{-0.0304}  & \textbf{0.0096}  & \textbf{0.3097}  & \textbf{160.3\%} \\
    \midrule
    \midrule
          & \multicolumn{18}{c}{KuaiRec-Large} \\
\cmidrule{2-19}    \textbf{Backbone} & \underline{0.7241}  & \underline{0.4021}  & 0.5732  & 0.1033  & 0.0000  &   -  & \underline{0.7120}  & \underline{0.3900}  & \underline{0.5834}  & \underline{0.1020}  & 0.0000  &   -  & \underline{0.6956}  & 0.3382  & 0.4679  & 0.0762  & \underline{0.0062}  & -\\
    \textbf{PD} & 0.6478  & 0.3388  & 0.5917  & 0.1062  & 0.0000  & 93.5\% & 0.5648  & 0.2644  & 0.7003  & 0.1241  & 0.0000  & 81.6\% & 0.6114  & 0.2961  & 0.6017  & 0.0994  & 0.0000  & 80.9\% \\
    \textbf{Pearson} & 0.6041  & 0.3079  & 0.5945  & 0.1054  & 0.0000  & 90.7\% & 0.5642  & 0.2630  & 0.7043  & 0.1249  & 0.0000  & 81.3\% & 0.6515  & 0.3129  & 0.5334  & 0.0812  & 0.0000  & 90.2\% \\
    \textbf{CNIF} & 0.6638  & 0.3477  & \underline{0.4789}  & \underline{0.0836}  & 0.0000  & \underline{103.9\%} & 0.5588  & 0.2890  & 0.6480  & 0.1231  & 0.0000  & \underline{85.3\%} & 0.6877  & \underline{0.3408}  & \underline{0.4232}  & \underline{0.0692}  & 0.0000  & \underline{104.2\%} \\
    \textbf{\sysname} & \textbf{0.8252}  & \textbf{0.4280}  & \textbf{-0.0547}  & \textbf{0.0224}  & \textbf{0.3440}  & \textbf{161.1\%} & \textbf{0.8445}  & \textbf{0.4609}  & \textbf{-0.0377}  & \textbf{0.0251}  & \textbf{0.3168}  & \textbf{163.9\%} & \textbf{0.7365}  & \textbf{0.3530}  & \textbf{0.0184}  & \textbf{0.0277}  & \textbf{0.2708}  & \textbf{149.0\%} \\
    \midrule
    \midrule
          & \multicolumn{18}{c}{Steam} \\
\cmidrule{2-19}    \textbf{Backbone} & \textbf{0.5328}  & \textbf{0.2882}  & 0.7401  & 0.0669  & 0.0000  &   -   & \textbf{0.5258}  & \textbf{0.2835}  & 0.7278  & 0.0621  & 0.0000  &  -   & \textbf{0.5049}  & \textbf{0.2518}  & 0.6540  & 0.0454  & \underline{0.0788}  & -\\
    \textbf{PD} & 0.4966  & 0.2441  & 0.7431  & 0.0652  & 0.0000  & \underline{96.5\%} & 0.4884  & 0.2511  & 0.7666  & 0.0673  & 0.0000  & 94.0\% & 0.4140  & 0.1914  & 0.6334  & 0.0400  & 0.0009  & 93.2\% \\
    \textbf{Pearson} & 0.4929  & 0.2385  & 0.7471  & 0.0657  & 0.0000  & 95.9\% & 0.4905  & 0.2528  & 0.7686  & 0.0676  & 0.0000  & 94.0\% & 0.3719  & 0.1545  & 0.5178  & \underline{0.0339}  & 0.0374  & 97.6\% \\
    \textbf{CNIF} & 0.3485  & 0.1713  & \underline{0.5537}  & \underline{0.0453}  & 0.0000  & 96.1\% & \underline{0.5137}  & \underline{0.2762}  & \underline{0.7095}  & \underline{0.0608}  & 0.0000  & \underline{100.1\%} & 0.4031  & 0.2071  & \underline{0.4658}  & 0.0453  & 0.0528  & \underline{103.9\%} \\
    \textbf{\sysname} & \underline{0.5143} & \underline{0.2674} &\textbf{0.4584} &\textbf{0.0325} &\textbf{0.1348} &\textbf{116.8\%}
 & 0.4458  & 0.2129  & \textbf{0.3905}  & \textbf{0.0370}  & \textbf{0.1597}  & \textbf{114.1\%} & \underline{0.4602}  & \underline{0.2151}  & \textbf{0.3794}  & \textbf{0.0321}  & \textbf{0.1391}  & \textbf{115.6\%} \\
    \bottomrule
    \end{tabular}%
}
\end{table*}

\begin{figure}[t]
\setlength{\abovecaptionskip}{6pt}
\centerline{
\includegraphics[width=0.4\textwidth]{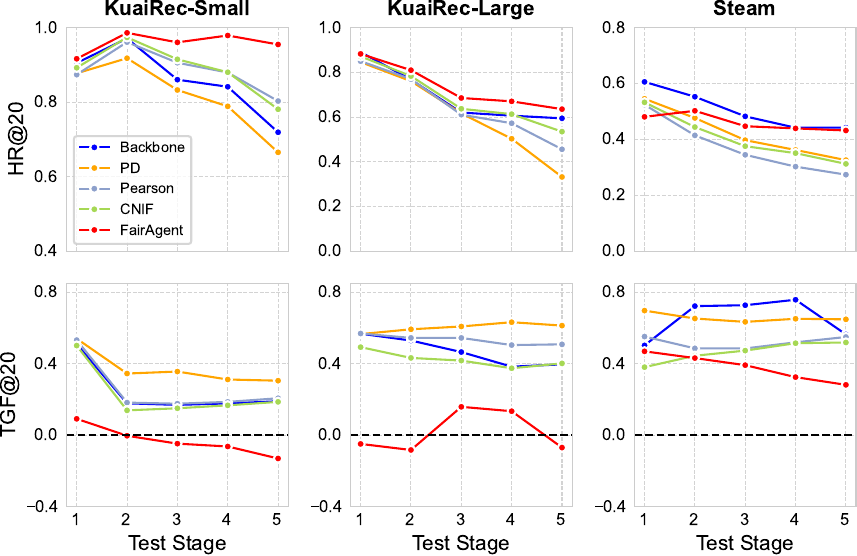}}
\caption{Dynamic changes in HR and TGF metrics across 5 test stages for different methods using the ALDI backbone.}
\label{fig:result_rq1_aldi}
\vspace{-0.4cm}
\end{figure}

We present how recommendation accuracy and new-item fairness of different methods change along with five stages, across different datasets in a dynamic recommendation scenario. 
Fig~\ref{fig:result_rq1_mf} and Fig~\ref{fig:result_rq1_aldi} present the results with MF and ALDI as the backbone model respectively\footnote{The trend for LightGCN is similar to MF and is omitted here due to space constraints.}, with sub-figures in a column representing the results under a dataset (3 in total in each figure).
In both the figures, the first row represents recommendation accuracy ($HR$, the higher the better), while the second row represents the degree of unfairness on new-item exposure ($TGF$, the closer to $0$ the better).
We also present some quantitative results obtained in the same experiments in Tab~\ref{tab:average_results},  where each value represents the average performance of the corresponding method across the five test stages, providing a measure of its overall effectiveness in DRS settings.
Bold values indicate the highest in each column, while underlined values represent the second highest. We make the following observations. 

\emph{First, \sysname can be effectively applied to different types of recommendation backbone models to increase exposure rate of new-items.}
From the analysis of NC (new-item coverage) metric (Tab~\ref{tab:average_results}), we found that with all backbones, the existing methods fail to improve the exposure rate of new-items (NC persistently remains near 0). 
These results highlight their inability to handle the continuous introduction of new-items in DRSs. 
In contrast, \sysname consistently achieves significant improvements in NC across all the backbone models and datasets (with NC approaching 30\% on KuaiRec-Small and KuaiRec-Large, and exceeding 13\% on Steam). This demonstrates its robust ability to enhance new-item exposure and effectively tackle the challenge of continuously introducing new-items in DRSs.

\emph{Second, \sysname is the most effective in improving the exposure fairness of new-items against old-items. It almost always achieves the lowest values of TGF under different backbone models and datasets}, while the other baselines typically show much higher TGF values (more serious unfairness), as observed from Fig~\ref{fig:result_rq1_mf}, Fig~\ref{fig:result_rq1_aldi} and Tab~\ref{tab:average_results}. 
A TGF value closer to 0 indicates smaller exposure disparities between new and old items, reflecting greater fairness in the DRS. 
For KuaiRec-Small and KuaiRec-Large, \sysname effectively captures users' strong preferences for new-items, significantly improving fairness. 
Specifically, its TGF values reaches close to $0$, and decrease by an average of $86.34\%$ and $93.35\%$ compared to all the backbone models on the two datasets, respectively. 
In contrast, the best baseline, CNIF, only achieves an decremental of $5.56\%$ and $4.98\%$.
For Steam, \sysname adopts a more balanced fairness optimization strategy to aligning with users' weaker preferences for new-items. 
Under this condition, TGF still improves by an average of $42.13\%$ compared to the backbone models, outperforming CNIF's $18.83\%$ improvement.

\emph{Third, \sysname maintains the highest recommendation accuracy in most of the cases, highlighting its ability in well balancing new-item fairness with accuracy in DRSs.}
Specifically, for KuaiRec-Small, \sysname maintains the highest recommendation accuracy across all the test stages, accurately recommending preferred new-items for users in dynamic recommendation. 
Compared to all the backbone models, the HR and NDCG metrics got improved by an average of $39.41\%$ and $61.42\%$, respectively.
In contrast, to achieve fairness, the best baseline CNIF sacrifices $2.19\%$ in HR.
For KuaiRec-Large, a similar pattern can be observed. 
Compared to the backbone models, \sysname achieves an improvement of $12.82\%$ in HR and $9.66\%$ in NDCG on average, while the best baseline suffers the losses of $10.33\%$ and $12.88\%$, respectively.
In Steam-like scenarios, where users have weaker interest in new-items, improving fairness often reduces recommendation accuracy by limiting old-item exposure.
This especially requires balancing well fairness and accuracy, as neither severe unfairness nor low accuracy suits practical needs.
We use $\delta T$ metric to evaluate how well different methods balances these two.
The results shows that\emph{\sysname achieves the best balance in all the scenarios, enhancing fairness with minimal loss in accuracy, thus maintaining long-term stability of a system.}


The analysis of the dataset and backbone model reveals several key insights. First, datasets where new-items and user preference evolve more rapidly seem to be more challenging. For instance, the accuracy and fairness of the backbone models decrease relatively more sharply with KuaiRec-Small, where user interest tend to quickly shift towards the numerous new-items entering overtime.  This confirms our concern raised in Section~\ref{section:fairness_concern}  that if unfairness is not addressed promptly, it may quickly accumulate in the dynamic feedback of DRSs, leading to reduced exposure of new-items and a sharp decline in recommendation accuracy.
And for the fairness enhancement methods, it seems more difficult to keep up with the dynamics in KuaiRec-Small.
For KuaiRec-Large and especially Steam, where the expansion of new-item set and also shift of user interest happen more slowly, the backbone models perform slightly better and more stably. The same goes for the baseline fairness methods. 
Second, despite being a cold-start model, ALDI does not consistently enhance the new-item exposure and fairness as expected.
Third, TGF of \sysname has negative values sometimes, meaning the reversal of exposure resources. 
This improves the chances of users encountering new-items. 
And its degree can be controlled to near zero through the use of adaptable parameters in the reward function. 




In summary, we answer RQ1: 
\textbf{{\emph{\sysname most effectively improves new-item exposure and enhances the fairness between new and old items, while maintaining high recommendation accuracy in DRSs.}}}

\begin{figure}[t]
    \setlength{\abovecaptionskip}{6pt}   
    \centering
    \includegraphics[width=0.45\textwidth]{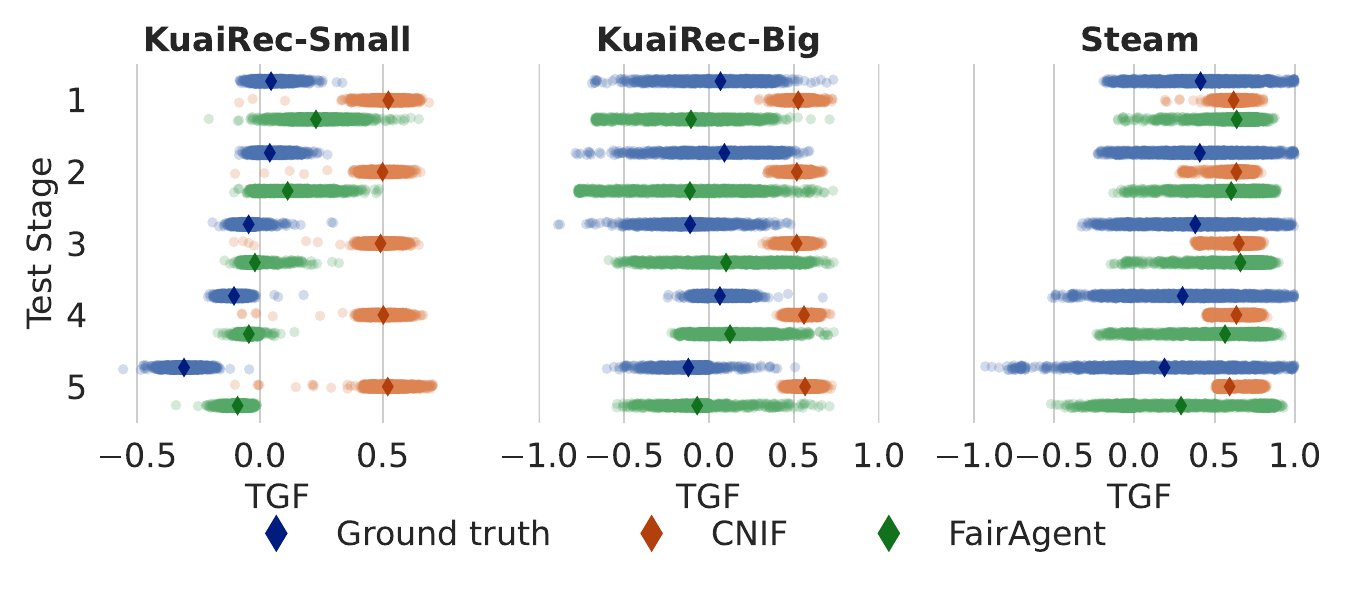}
            \caption{Performance of \sysname and the best baseline in adapting to users' true preferences for new-items across different recommendation stages.}
\label{fig:user_dynamic}
\vspace{-0.45cm}
\end{figure}

\subsubsection{Results of RQ2}
Tab~\ref{tab:average_results} presents the average UNF (user-level new-item fairness) values across the five test stages, with smaller values indicating that the recommendations better adapting to the dynamic changes in users' personalized preferences for new-items. 
The results show that \sysname consistently achieves the best UNF in all the scenarios. 
To illustrate this more intuitively, we visualize the phenomenon in Fig~\ref{fig:user_dynamic}. 
Each point in the blue area represents a user's true preference for new-items at a specific recommendation stage, with the value calculated using the TGF metric based on ground-truth interactions. 
The orange and green areas represent the recommendation results of CNIF\footnote{CNIF is the best performed baseline for improving new-item fairness, so we choose it for comparison. Since the conclusions are similar across different backbone models, we present the results for MF only due to space constraints.}, and \sysname, respectively, with values calculated using the TGF metric based on their respective recommendation lists. 
When the distribution along the x-axis closely matches the blue region (with bars of equal length) in the same test stage, it signifies that the recommendation results better align with the user's dynamic fairness requirements.

From the results on KuaiRec-Small, we observe that most users' preferences shift with the introduction of new-items (TGF shifts from near zero to negative), and \sysname effectively tracks these changes (green areas closely follow the movement of blue areas).
This shows that \sysname captures the dynamic evolution of users' personalized preferences for new-items and adjusts its fairness strategies accordingly.
In contrast, CNIF fails to do so, leading to significant mismatches between recommendations and users' actual preferences.
On KuaiRec-Large, users show diverse preferences for new-items, with some favoring them (TGF < 0) and others preferring older ones (TGF > 0).
\sysname adapts to this diversity by tailoring fairness strategies to individual users, demonstrating its ability to achieve new-item fairness at the user level in DRSs.
In the Steam scenario, most users prefer older items (TGF > 0), and \sysname successfully captures this.
It adopts a milder fairness strategy that not only improves new-item fairness but also respects users' preferences for older items, striking an effective balance between accuracy and fairness.

In summary, we answer RQ2: {\emph{\textbf{\sysname can effectively adapt to users' personalized dynamic changes in preferences for old and new items within DRSs by flexibly tailoring its optimization strategies.}}}

\subsubsection{Results on RQ3.} 
\begin{table}[tp]
\setlength{\abovecaptionskip}{3pt}
\setlength\tabcolsep{3pt}
\centering
\footnotesize
\caption{Results of ablation study on three datasets.}
\vspace{-1pt}
\resizebox{0.8\linewidth}{!}{
\begin{tabular}{cccccc}
\toprule
\multirow{2}[3]{*}{\textbf{Method}} & \multicolumn{5}{c}{\textbf{Dataset - KuaiRec-Small}} \\
\cmidrule{2-6}      & HR$\uparrow$    & NDCG$\uparrow$  & TGF$\downarrow$   & UNF$\downarrow$   & NC$\uparrow$ \\
\midrule
\textbf{\sysname} & \textbf{0.9600 } & \textbf{0.5082 } & \textbf{0.0793 } & \textbf{0.0061 } & \textbf{0.3003 } \\
\textbf{$\sysname_{w/o\ f}$} & \underline{0.9340}  & \underline{0.4617}  & 0.1985  & 0.0576  & \underline{0.1945}   \\
\textbf{$\sysname_{w/o\ n}$} & 0.9315  & 0.4605   & \underline{0.1824}  & \underline{0.0370}  & 0.1699  \\
\textbf{$\sysname_{w/o\ f\&n}$} & 0.7946  & 0.3461  & 0.4649  & 0.0410  & 0.0450  \\
\midrule
\midrule
      & \multicolumn{5}{c}{\textbf{Dataset - KuaiRec-Large}} \\
\midrule
\textbf{\sysname} & \underline{0.8445}& \underline{0.4609} & \textbf{-0.0377} & \textbf{0.0251} & \underline{0.3168} \\
\textbf{$\sysname_{w/o\ f}$} &  \textbf{0.8686} & \textbf{0.4897} &0.0786 & 0.0312 & \textbf{0.3562}\\
\textbf{$\sysname_{w/o\ n}$} & 0.8311 & 0.4537 &\underline{-0.0685} & \underline{0.0270} & 0.2704\\
\textbf{$\sysname_{w/o\ f\&n}$} & 0.7574 & 0.4052 & 0.3161 &0.0513 &0.1343\\
\midrule
\midrule
      & \multicolumn{5}{c}{\textbf{Dataset - Steam}} \\
      \midrule
\textbf{\sysname} & 0.4477 & 0.2369 & \underline{0.4438} & \textbf{0.0250} & \underline{0.1733}\\
\textbf{$\sysname_{w/o\ f}$} & 0.4055 & 0.1768  & 0.4584 & 0.0396 & \textbf{0.1918}\\
\textbf{$\sysname_{w/o\ n}$} & \textbf{0.5143} & \textbf{0.2674} & \textbf{0.4051} & \underline{0.0325} & 0.1348\\
\textbf{$\sysname_{w/o\ f\&n}$} & \underline{0.5028} & \underline{0.2593} & 0.5401 & 0.0425 & 0.0938 \\
\bottomrule
\end{tabular}%
}
\vspace{-0.2cm}
\label{tab:ablation_study}
\end{table}

To analyze the roles of different reward components, we conducted ablation studies. 
As similar conclusions were drawn across three backbone models, we present only the MF results in Tab~\ref{tab:ablation_study}. 
Specifically, $\sysname_{w/o\ f}$ removes the fairness reward $R_{fair}$ (let $\alpha = 0$), $\sysname_{w/o\ n}$ removes the new-item exploration reward $R_{new}$ (let $\beta = 0$), and $\sysname_{w/o\ f\&n}$ removes both (let $\alpha = 0, \beta = 0$).
We observe that removing \( R_{fair} \) leads to poorer fairness performance, as indicated by increased \(|\text{TGF}|\) and UNF values. This validates the effectiveness of the fairness reward in improving new-item fairness in DRSs by considering users' personalized preferences for new-items. 
Similarly, removing \( R_{new} \) results in a significant drop in new-item recommendation rates, as reflected by the reduced NC values. 
This demonstrates the effectiveness of \( R_{new} \) in enhancing the exposure of dynamically introduced new-items.
In scenarios where users prefer new-items (e.g., KuaiRec-Small), $R_{fair}$ and $R_{new}$ can work synergistically to both ensure new-item fairness and significantly enhance recommendation accuracy.

In summary, we answer RQ3: \emph{\textbf{Our designed fairness reward and new-item exploration reward effectively enhance new-item fairness and increase new-item recommendation rates.}}

\enlargethispage{0.2cm}
\section{Conclusion and Discussion}
In this work, we proposed \sysname, a RL-based new-item fairness enhancement framework, designed to effectively boost new-item exposure, keep up with users' personalized preferences for new-items, and meanwhile maintain high recommendation accuracy within DRSs. 

While \sysname shows promising performance, several directions remain for future improvement. 
First, its reliance on a backbone model may constrain performance due to the model’s inherent limitations, whereas RL-based approaches often suffer from slow convergence and poor scalability. 
Enhancing computational efficiency would also improve applicability in large-scale settings. 
Finally, extending fairness considerations to other stakeholders, such as content creators and platform providers, could lead to more comprehensive and equitable DRSs.

\section{Acknowledgment}
This research is supported by the State Key Laboratory of Industrial Control Technology, China (Grant No. ICT2024C01), and partially supported by the Ministry of Education, Singapore, under its MOE AcRF Tier 1 SUTD Kickstarter Initiative (SKI 2021\_06\_12) and MOE AcRF Tier 1 Grant (RG13/23).

\newpage
\clearpage
\balance
\bibliographystyle{ACM-Reference-Format}
\bibliography{ref} 


\begin{thebibliography}{43}


\ifx \showCODEN    \undefined \def \showCODEN     #1{\unskip}     \fi
\ifx \showDOI      \undefined \def \showDOI       #1{#1}\fi
\ifx \showISBNx    \undefined \def \showISBNx     #1{\unskip}     \fi
\ifx \showISBNxiii \undefined \def \showISBNxiii  #1{\unskip}     \fi
\ifx \showISSN     \undefined \def \showISSN      #1{\unskip}     \fi
\ifx \showLCCN     \undefined \def \showLCCN      #1{\unskip}     \fi
\ifx \shownote     \undefined \def \shownote      #1{#1}          \fi
\ifx \showarticletitle \undefined \def \showarticletitle #1{#1}   \fi
\ifx \showURL      \undefined \def \showURL       {\relax}        \fi
\providecommand\bibfield[2]{#2}
\providecommand\bibinfo[2]{#2}
\providecommand\natexlab[1]{#1}
\providecommand\showeprint[2][]{arXiv:#2}

\bibitem[Agarwal et~al\mbox{.}(2019)]%
        {agarwal2019estimating}
\bibfield{author}{\bibinfo{person}{Aman Agarwal}, \bibinfo{person}{Ivan Zaitsev}, \bibinfo{person}{Xuanhui Wang}, \bibinfo{person}{Cheng Li}, \bibinfo{person}{Marc Najork}, {and} \bibinfo{person}{Thorsten Joachims}.} \bibinfo{year}{2019}\natexlab{}.
\newblock \showarticletitle{Estimating position bias without intrusive interventions}. In \bibinfo{booktitle}{\emph{Proceedings of the twelfth ACM International Conference on Web Search and Data Mining}}. \bibinfo{pages}{474--482}.
\newblock


\bibitem[Chen et~al\mbox{.}(2022)]%
        {chen2022generative}
\bibfield{author}{\bibinfo{person}{Hao Chen}, \bibinfo{person}{Zefan Wang}, \bibinfo{person}{Feiran Huang}, \bibinfo{person}{Xiao Huang}, \bibinfo{person}{Yue Xu}, \bibinfo{person}{Yishi Lin}, \bibinfo{person}{Peng He}, {and} \bibinfo{person}{Zhoujun Li}.} \bibinfo{year}{2022}\natexlab{}.
\newblock \showarticletitle{Generative adversarial framework for cold-start item recommendation}. In \bibinfo{booktitle}{\emph{Proceedings of the 45th International ACM SIGIR Conference on Research and Development in Information Retrieval}}. \bibinfo{pages}{2565--2571}.
\newblock


\bibitem[Chen et~al\mbox{.}(2024)]%
        {chen2024opportunities}
\bibfield{author}{\bibinfo{person}{Xiaocong Chen}, \bibinfo{person}{Siyu Wang}, \bibinfo{person}{Julian McAuley}, \bibinfo{person}{Dietmar Jannach}, {and} \bibinfo{person}{Lina Yao}.} \bibinfo{year}{2024}\natexlab{}.
\newblock \showarticletitle{On the opportunities and challenges of offline reinforcement learning for recommender systems}.
\newblock \bibinfo{journal}{\emph{ACM Transactions on Information Systems}} \bibinfo{volume}{42}, \bibinfo{number}{6} (\bibinfo{year}{2024}), \bibinfo{pages}{1--26}.
\newblock


\bibitem[Du et~al\mbox{.}(2025)]%
        {du2025quasi}
\bibfield{author}{\bibinfo{person}{Yingpeng Du}, \bibinfo{person}{Hongzhi Liu}, \bibinfo{person}{Hengshu Zhu}, \bibinfo{person}{Yang Song}, \bibinfo{person}{Zhi Zheng}, {and} \bibinfo{person}{Zhonghai Wu}.} \bibinfo{year}{2025}\natexlab{}.
\newblock \showarticletitle{Quasi-Metric Learning for Bilateral Person-Job Fit}.
\newblock \bibinfo{journal}{\emph{IEEE Transactions on Pattern Analysis and Machine Intelligence}} (\bibinfo{year}{2025}).
\newblock


\bibitem[Du et~al\mbox{.}(2024a)]%
        {du2024enhancing}
\bibfield{author}{\bibinfo{person}{Yingpeng Du}, \bibinfo{person}{Di Luo}, \bibinfo{person}{Rui Yan}, \bibinfo{person}{Xiaopei Wang}, \bibinfo{person}{Hongzhi Liu}, \bibinfo{person}{Hengshu Zhu}, \bibinfo{person}{Yang Song}, {and} \bibinfo{person}{Jie Zhang}.} \bibinfo{year}{2024}\natexlab{a}.
\newblock \showarticletitle{Enhancing job recommendation through llm-based generative adversarial networks}. In \bibinfo{booktitle}{\emph{Proceedings of the AAAI Conference on Artificial Intelligence}}, Vol.~\bibinfo{volume}{38}. \bibinfo{pages}{8363--8371}.
\newblock


\bibitem[Du et~al\mbox{.}(2024b)]%
        {du2024disentangled}
\bibfield{author}{\bibinfo{person}{Yingpeng Du}, \bibinfo{person}{Ziyan Wang}, \bibinfo{person}{Zhu Sun}, \bibinfo{person}{Yining Ma}, \bibinfo{person}{Hongzhi Liu}, {and} \bibinfo{person}{Jie Zhang}.} \bibinfo{year}{2024}\natexlab{b}.
\newblock \showarticletitle{Disentangled Multi-interest Representation Learning for Sequential Recommendation}. In \bibinfo{booktitle}{\emph{Proceedings of the 30th ACM SIGKDD Conference on Knowledge Discovery and Data Mining}}. \bibinfo{pages}{677--688}.
\newblock


\bibitem[Du et~al\mbox{.}(2022)]%
        {du2022metakg}
\bibfield{author}{\bibinfo{person}{Yuntao Du}, \bibinfo{person}{Xinjun Zhu}, \bibinfo{person}{Lu Chen}, \bibinfo{person}{Ziquan Fang}, {and} \bibinfo{person}{Yunjun Gao}.} \bibinfo{year}{2022}\natexlab{}.
\newblock \showarticletitle{Metakg: Meta-learning on knowledge graph for cold-start recommendation}.
\newblock \bibinfo{journal}{\emph{IEEE Transactions on knowledge and data engineering}} \bibinfo{volume}{35}, \bibinfo{number}{10} (\bibinfo{year}{2022}), \bibinfo{pages}{9850--9863}.
\newblock


\bibitem[Gao et~al\mbox{.}(2022)]%
        {gao2022kuairec}
\bibfield{author}{\bibinfo{person}{Chongming Gao}, \bibinfo{person}{Shijun Li}, \bibinfo{person}{Wenqiang Lei}, \bibinfo{person}{Jiawei Chen}, \bibinfo{person}{Biao Li}, \bibinfo{person}{Peng Jiang}, \bibinfo{person}{Xiangnan He}, \bibinfo{person}{Jiaxin Mao}, {and} \bibinfo{person}{Tat-Seng Chua}.} \bibinfo{year}{2022}\natexlab{}.
\newblock \showarticletitle{KuaiRec: A Fully-Observed Dataset and Insights for Evaluating Recommender Systems}. In \bibinfo{booktitle}{\emph{Proceedings of the 31st ACM International Conference on Information \& Knowledge Management}} (Atlanta, GA, USA) \emph{(\bibinfo{series}{CIKM '22})}. \bibinfo{pages}{540–550}.
\newblock
\urldef\tempurl%
\url{https://doi.org/10.1145/3511808.3557220}
\showDOI{\tempurl}


\bibitem[Ge et~al\mbox{.}(2024)]%
        {ge2024short}
\bibfield{author}{\bibinfo{person}{Shiping Ge}, \bibinfo{person}{Qiang Chen}, \bibinfo{person}{Zhiwei Jiang}, \bibinfo{person}{Yafeng Yin}, \bibinfo{person}{Ziyao Chen}, {and} \bibinfo{person}{Qing Gu}.} \bibinfo{year}{2024}\natexlab{}.
\newblock \showarticletitle{Short Video Ordering via Position Decoding and Successor Prediction}. In \bibinfo{booktitle}{\emph{Proceedings of the 47th International ACM SIGIR Conference on Research and Development in Information Retrieval}}. \bibinfo{pages}{2167--2176}.
\newblock


\bibitem[Guo et~al\mbox{.}(2024)]%
        {guo2024configurable}
\bibfield{author}{\bibinfo{person}{Huizhong Guo}, \bibinfo{person}{Dongxia Wang}, \bibinfo{person}{Zhu Sun}, \bibinfo{person}{Haonan Zhang}, \bibinfo{person}{Jinfeng Li}, {and} \bibinfo{person}{Jie Zhang}.} \bibinfo{year}{2024}\natexlab{}.
\newblock \showarticletitle{Configurable Fairness for New Item Recommendation Considering Entry Time of Items}. In \bibinfo{booktitle}{\emph{Proceedings of the 47th International ACM SIGIR Conference on Research and Development in Information Retrieval}}. \bibinfo{pages}{437--447}.
\newblock


\bibitem[He et~al\mbox{.}(2020)]%
        {he2020lightgcn}
\bibfield{author}{\bibinfo{person}{Xiangnan He}, \bibinfo{person}{Kuan Deng}, \bibinfo{person}{Xiang Wang}, \bibinfo{person}{Yan Li}, \bibinfo{person}{Yongdong Zhang}, {and} \bibinfo{person}{Meng Wang}.} \bibinfo{year}{2020}\natexlab{}.
\newblock \showarticletitle{Lightgcn: Simplifying and powering graph convolution network for recommendation}. In \bibinfo{booktitle}{\emph{Proceedings of the 43rd International ACM SIGIR Conference on Research and Development in Information Retrieval}}. \bibinfo{pages}{639--648}.
\newblock


\bibitem[Huang et~al\mbox{.}(2023)]%
        {Huang2023aligning}
\bibfield{author}{\bibinfo{person}{Feiran Huang}, \bibinfo{person}{Zefan Wang}, \bibinfo{person}{Xiao Huang}, \bibinfo{person}{Yufeng Qian}, \bibinfo{person}{Zhetao Li}, {and} \bibinfo{person}{Hao Chen}.} \bibinfo{year}{2023}\natexlab{}.
\newblock \showarticletitle{Aligning Distillation For Cold-Start Item Recommendation}. In \bibinfo{booktitle}{\emph{Proceedings of the 46th International ACM SIGIR Conference on Research and Development in Information Retrieval}}. \bibinfo{pages}{1147–1157}.
\newblock
\urldef\tempurl%
\url{https://doi.org/10.1145/3539618.3591732}
\showDOI{\tempurl}


\bibitem[J{\"a}rvelin and Kek{\"a}l{\"a}inen(2002)]%
        {jarvelin2002cumulated}
\bibfield{author}{\bibinfo{person}{Kalervo J{\"a}rvelin} {and} \bibinfo{person}{Jaana Kek{\"a}l{\"a}inen}.} \bibinfo{year}{2002}\natexlab{}.
\newblock \showarticletitle{Cumulated gain-based evaluation of IR techniques}.
\newblock \bibinfo{journal}{\emph{ACM Transactions on Information Systems (TOIS)}} \bibinfo{volume}{20}, \bibinfo{number}{4} (\bibinfo{year}{2002}), \bibinfo{pages}{422--446}.
\newblock


\bibitem[Kim et~al\mbox{.}(2024)]%
        {kim2024content}
\bibfield{author}{\bibinfo{person}{Jinri Kim}, \bibinfo{person}{Eungi Kim}, \bibinfo{person}{Kwangeun Yeo}, \bibinfo{person}{Yujin Jeon}, \bibinfo{person}{Chanwoo Kim}, \bibinfo{person}{Sewon Lee}, {and} \bibinfo{person}{Joonseok Lee}.} \bibinfo{year}{2024}\natexlab{}.
\newblock \showarticletitle{Content-based graph reconstruction for cold-start item recommendation}. In \bibinfo{booktitle}{\emph{Proceedings of the 47th International ACM SIGIR Conference on Research and Development in Information Retrieval}}. \bibinfo{pages}{1263--1273}.
\newblock


\bibitem[Liu et~al\mbox{.}(2021)]%
        {liu2021top}
\bibfield{author}{\bibinfo{person}{Hongyang Liu}, \bibinfo{person}{Zhu Sun}, \bibinfo{person}{Xinghua Qu}, {and} \bibinfo{person}{Fuyong Yuan}.} \bibinfo{year}{2021}\natexlab{}.
\newblock \showarticletitle{Top-aware recommender distillation with deep reinforcement learning}.
\newblock \bibinfo{journal}{\emph{Information Sciences}}  \bibinfo{volume}{576} (\bibinfo{year}{2021}), \bibinfo{pages}{642--657}.
\newblock


\bibitem[Liu et~al\mbox{.}(2023)]%
        {liu2023mitigating}
\bibfield{author}{\bibinfo{person}{Zhongzhou Liu}, \bibinfo{person}{Yuan Fang}, {and} \bibinfo{person}{Min Wu}.} \bibinfo{year}{2023}\natexlab{}.
\newblock \showarticletitle{Mitigating popularity bias for users and items with fairness-centric adaptive recommendation}.
\newblock \bibinfo{journal}{\emph{ACM Transactions on Information Systems}} \bibinfo{volume}{41}, \bibinfo{number}{3} (\bibinfo{year}{2023}), \bibinfo{pages}{1--27}.
\newblock


\bibitem[Mnih(2013)]%
        {mnih2013playing}
\bibfield{author}{\bibinfo{person}{Volodymyr Mnih}.} \bibinfo{year}{2013}\natexlab{}.
\newblock \showarticletitle{Playing atari with deep reinforcement learning}.
\newblock \bibinfo{journal}{\emph{arXiv preprint arXiv:1312.5602}} (\bibinfo{year}{2013}).
\newblock


\bibitem[Morik et~al\mbox{.}(2020)]%
        {morik2020controlling}
\bibfield{author}{\bibinfo{person}{Marco Morik}, \bibinfo{person}{Ashudeep Singh}, \bibinfo{person}{Jessica Hong}, {and} \bibinfo{person}{Thorsten Joachims}.} \bibinfo{year}{2020}\natexlab{}.
\newblock \showarticletitle{Controlling fairness and bias in dynamic learning-to-rank}. In \bibinfo{booktitle}{\emph{Proceedings of the 43rd international ACM SIGIR Conference on Research and Development in Information Retrieval}}. \bibinfo{pages}{429--438}.
\newblock


\bibitem[O'Neill et~al\mbox{.}(2016)]%
        {o2016condensing}
\bibfield{author}{\bibinfo{person}{Mark O'Neill}, \bibinfo{person}{Elham Vaziripour}, \bibinfo{person}{Justin Wu}, {and} \bibinfo{person}{Daniel Zappala}.} \bibinfo{year}{2016}\natexlab{}.
\newblock \showarticletitle{Condensing steam: Distilling the diversity of gamer behavior}. In \bibinfo{booktitle}{\emph{Proceedings of the 2016 Internet Measurement Conference}}. \bibinfo{pages}{81--95}.
\newblock


\bibitem[O’Neill et~al\mbox{.}(2010)]%
        {o2010play}
\bibfield{author}{\bibinfo{person}{Joseph O’Neill}, \bibinfo{person}{Barty Pleydell-Bouverie}, \bibinfo{person}{David Dupret}, {and} \bibinfo{person}{Jozsef Csicsvari}.} \bibinfo{year}{2010}\natexlab{}.
\newblock \showarticletitle{Play it again: reactivation of waking experience and memory}.
\newblock \bibinfo{journal}{\emph{Trends in Neurosciences}} \bibinfo{volume}{33}, \bibinfo{number}{5} (\bibinfo{year}{2010}), \bibinfo{pages}{220--229}.
\newblock


\bibitem[Reimers and Gurevych(2019)]%
        {reimers-2019-sentence-bert}
\bibfield{author}{\bibinfo{person}{Nils Reimers} {and} \bibinfo{person}{Iryna Gurevych}.} \bibinfo{year}{2019}\natexlab{}.
\newblock \showarticletitle{Sentence-BERT: Sentence Embeddings using Siamese BERT-Networks}. In \bibinfo{booktitle}{\emph{Proceedings of the 2019 Conference on Empirical Methods in Natural Language Processing}}. \bibinfo{publisher}{Association for Computational Linguistics}.
\newblock
\urldef\tempurl%
\url{http://arxiv.org/abs/1908.10084}
\showURL{%
\tempurl}


\bibitem[Rendle et~al\mbox{.}(2012)]%
        {rendle2012bpr}
\bibfield{author}{\bibinfo{person}{Steffen Rendle}, \bibinfo{person}{Christoph Freudenthaler}, \bibinfo{person}{Zeno Gantner}, {and} \bibinfo{person}{Lars Schmidt-Thieme}.} \bibinfo{year}{2012}\natexlab{}.
\newblock \showarticletitle{BPR: Bayesian personalized ranking from implicit feedback}.
\newblock \bibinfo{journal}{\emph{arXiv preprint arXiv:1205.2618}} (\bibinfo{year}{2012}).
\newblock


\bibitem[Rhee et~al\mbox{.}(2022)]%
        {rhee2022countering}
\bibfield{author}{\bibinfo{person}{Wondo Rhee}, \bibinfo{person}{Sung~Min Cho}, {and} \bibinfo{person}{Bongwon Suh}.} \bibinfo{year}{2022}\natexlab{}.
\newblock \showarticletitle{Countering Popularity Bias by Regularizing Score Differences}. In \bibinfo{booktitle}{\emph{Proceedings of the 16th ACM Conference on Recommender Systems}}. \bibinfo{pages}{145--155}.
\newblock


\bibitem[Schaul(2015)]%
        {schaul2015prioritized}
\bibfield{author}{\bibinfo{person}{Tom Schaul}.} \bibinfo{year}{2015}\natexlab{}.
\newblock \showarticletitle{Prioritized Experience Replay}.
\newblock \bibinfo{journal}{\emph{arXiv preprint arXiv:1511.05952}} (\bibinfo{year}{2015}).
\newblock


\bibitem[Schnabel et~al\mbox{.}(2016)]%
        {schnabel2016recommendations}
\bibfield{author}{\bibinfo{person}{Tobias Schnabel}, \bibinfo{person}{Adith Swaminathan}, \bibinfo{person}{Ashudeep Singh}, \bibinfo{person}{Navin Chandak}, {and} \bibinfo{person}{Thorsten Joachims}.} \bibinfo{year}{2016}\natexlab{}.
\newblock \showarticletitle{Recommendations as treatments: Debiasing learning and evaluation}. In \bibinfo{booktitle}{\emph{International Conference on Machine Learning}}. PMLR, \bibinfo{pages}{1670--1679}.
\newblock


\bibitem[Sun et~al\mbox{.}(2022)]%
        {sun2022daisyrec}
\bibfield{author}{\bibinfo{person}{Zhu Sun}, \bibinfo{person}{Hui Fang}, \bibinfo{person}{Jie Yang}, \bibinfo{person}{Xinghua Qu}, \bibinfo{person}{Hongyang Liu}, \bibinfo{person}{Di Yu}, \bibinfo{person}{Yew-Soon Ong}, {and} \bibinfo{person}{Jie Zhang}.} \bibinfo{year}{2022}\natexlab{}.
\newblock \showarticletitle{DaisyRec 2.0: Benchmarking Recommendation for Rigorous Evaluation}.
\newblock \bibinfo{journal}{\emph{IEEE Transactions on Pattern Analysis and Machine Intelligence (TPAMI)}} (\bibinfo{year}{2022}).
\newblock


\bibitem[Sun et~al\mbox{.}(2020)]%
        {sun2020are}
\bibfield{author}{\bibinfo{person}{Zhu Sun}, \bibinfo{person}{Di Yu}, \bibinfo{person}{Hui Fang}, \bibinfo{person}{Jie Yang}, \bibinfo{person}{Xinghua Qu}, \bibinfo{person}{Jie Zhang}, {and} \bibinfo{person}{Cong Geng}.} \bibinfo{year}{2020}\natexlab{}.
\newblock \showarticletitle{Are We Evaluating Rigorously? Benchmarking Recommendation for Reproducible Evaluation and Fair Comparison}. In \bibinfo{booktitle}{\emph{Proceedings of the 14th ACM Conference on Recommender Systems}}.
\newblock


\bibitem[Volkovs et~al\mbox{.}(2017)]%
        {volkovs2017dropoutnet}
\bibfield{author}{\bibinfo{person}{Maksims Volkovs}, \bibinfo{person}{Guangwei Yu}, {and} \bibinfo{person}{Tomi Poutanen}.} \bibinfo{year}{2017}\natexlab{}.
\newblock \showarticletitle{Dropoutnet: Addressing cold start in recommender systems}.
\newblock \bibinfo{journal}{\emph{Advances in neural information processing systems}}  \bibinfo{volume}{30} (\bibinfo{year}{2017}).
\newblock


\bibitem[Wang et~al\mbox{.}(2021)]%
        {wang2021deconfounded}
\bibfield{author}{\bibinfo{person}{Wenjie Wang}, \bibinfo{person}{Fuli Feng}, \bibinfo{person}{Xiangnan He}, \bibinfo{person}{Xiang Wang}, {and} \bibinfo{person}{Tat-Seng Chua}.} \bibinfo{year}{2021}\natexlab{}.
\newblock \showarticletitle{Deconfounded recommendation for alleviating bias amplification}. In \bibinfo{booktitle}{\emph{Proceedings of the 27th ACM SIGKDD Conference on Knowledge Discovery \& Data Mining}}. \bibinfo{pages}{1717--1725}.
\newblock


\bibitem[Wang et~al\mbox{.}(2024a)]%
        {wang2024warming}
\bibfield{author}{\bibinfo{person}{Yaqing Wang}, \bibinfo{person}{Hongming Piao}, \bibinfo{person}{Daxiang Dong}, \bibinfo{person}{Quanming Yao}, {and} \bibinfo{person}{Jingbo Zhou}.} \bibinfo{year}{2024}\natexlab{a}.
\newblock \showarticletitle{Warming Up Cold-Start CTR Prediction by Learning Item-Specific Feature Interactions}. In \bibinfo{booktitle}{\emph{Proceedings of the 30th ACM SIGKDD Conference on Knowledge Discovery and Data Mining}}. \bibinfo{pages}{3233--3244}.
\newblock


\bibitem[Wang et~al\mbox{.}(2024b)]%
        {wang2024amazon}
\bibfield{author}{\bibinfo{person}{Yuhan Wang}, \bibinfo{person}{Qing Xie}, \bibinfo{person}{Mengzi Tang}, \bibinfo{person}{Lin Li}, \bibinfo{person}{Jingling Yuan}, {and} \bibinfo{person}{Yongjian Liu}.} \bibinfo{year}{2024}\natexlab{b}.
\newblock \showarticletitle{Amazon-KG: A Knowledge Graph Enhanced Cross-Domain Recommendation Dataset}. In \bibinfo{booktitle}{\emph{Proceedings of the 47th International ACM SIGIR Conference on Research and Development in Information Retrieval}}. \bibinfo{pages}{123--130}.
\newblock


\bibitem[Wei et~al\mbox{.}(2024)]%
        {wei2024fpsr+}
\bibfield{author}{\bibinfo{person}{Tianjun Wei}, \bibinfo{person}{Tommy~WS Chow}, {and} \bibinfo{person}{Jianghong Ma}.} \bibinfo{year}{2024}\natexlab{}.
\newblock \showarticletitle{FPSR+: Toward Robust, Efficient and Scalable Collaborative Filtering With Partition-aware Item Similarity Modeling}.
\newblock \bibinfo{journal}{\emph{IEEE Transactions on Knowledge and Data Engineering}} (\bibinfo{year}{2024}).
\newblock


\bibitem[Wei et~al\mbox{.}(2023)]%
        {wei2023collaborative}
\bibfield{author}{\bibinfo{person}{Tianjun Wei}, \bibinfo{person}{Jianghong Ma}, {and} \bibinfo{person}{Tommy~WS Chow}.} \bibinfo{year}{2023}\natexlab{}.
\newblock \showarticletitle{Collaborative residual metric learning}. In \bibinfo{booktitle}{\emph{Proceedings of the 46th International ACM SIGIR Conference on Research and Development in Information Retrieval}}. \bibinfo{pages}{1107--1116}.
\newblock


\bibitem[Xin et~al\mbox{.}(2020)]%
        {xin2020self}
\bibfield{author}{\bibinfo{person}{Xin Xin}, \bibinfo{person}{Alexandros Karatzoglou}, \bibinfo{person}{Ioannis Arapakis}, {and} \bibinfo{person}{Joemon~M Jose}.} \bibinfo{year}{2020}\natexlab{}.
\newblock \showarticletitle{Self-supervised reinforcement learning for recommender systems}. In \bibinfo{booktitle}{\emph{Proceedings of the 43rd International ACM SIGIR Conference on Research and Development in Information Retrieval}}. \bibinfo{pages}{931--940}.
\newblock


\bibitem[Yang et~al\mbox{.}(2023)]%
        {yang2023rectifying}
\bibfield{author}{\bibinfo{person}{Mengyue Yang}, \bibinfo{person}{Jun Wang}, {and} \bibinfo{person}{Jean-Francois Ton}.} \bibinfo{year}{2023}\natexlab{}.
\newblock \showarticletitle{Rectifying unfairness in recommendation feedback loop}. In \bibinfo{booktitle}{\emph{Proceedings of the 46th international ACM SIGIR Conference on Research and Development in Information Retrieval}}. \bibinfo{pages}{28--37}.
\newblock


\bibitem[Yoo et~al\mbox{.}(2024)]%
        {yoo2024ensuring}
\bibfield{author}{\bibinfo{person}{Hyunsik Yoo}, \bibinfo{person}{Zhichen Zeng}, \bibinfo{person}{Jian Kang}, \bibinfo{person}{Ruizhong Qiu}, \bibinfo{person}{David Zhou}, \bibinfo{person}{Zhining Liu}, \bibinfo{person}{Fei Wang}, \bibinfo{person}{Charlie Xu}, \bibinfo{person}{Eunice Chan}, {and} \bibinfo{person}{Hanghang Tong}.} \bibinfo{year}{2024}\natexlab{}.
\newblock \showarticletitle{Ensuring user-side fairness in dynamic recommender systems}. In \bibinfo{booktitle}{\emph{Proceedings of the ACM on Web Conference 2024}}. \bibinfo{pages}{3667--3678}.
\newblock


\bibitem[Zhang et~al\mbox{.}(2021)]%
        {zhang2021causal}
\bibfield{author}{\bibinfo{person}{Yang Zhang}, \bibinfo{person}{Fuli Feng}, \bibinfo{person}{Xiangnan He}, \bibinfo{person}{Tianxin Wei}, \bibinfo{person}{Chonggang Song}, \bibinfo{person}{Guohui Ling}, {and} \bibinfo{person}{Yongdong Zhang}.} \bibinfo{year}{2021}\natexlab{}.
\newblock \showarticletitle{Causal intervention for leveraging popularity bias in recommendation}. In \bibinfo{booktitle}{\emph{Proceedings of the 44th International ACM SIGIR Conference on Research and Development in Information Retrieval}}. \bibinfo{pages}{11--20}.
\newblock


\bibitem[Zhao et~al\mbox{.}(2018)]%
        {zhao2018recommendations}
\bibfield{author}{\bibinfo{person}{Xiangyu Zhao}, \bibinfo{person}{Liang Zhang}, \bibinfo{person}{Zhuoye Ding}, \bibinfo{person}{Long Xia}, \bibinfo{person}{Jiliang Tang}, {and} \bibinfo{person}{Dawei Yin}.} \bibinfo{year}{2018}\natexlab{}.
\newblock \showarticletitle{Recommendations with negative feedback via pairwise deep reinforcement learning}. In \bibinfo{booktitle}{\emph{Proceedings of the 24th ACM SIGKDD International Conference on Knowledge Discovery \& Data Mining}}. \bibinfo{pages}{1040--1048}.
\newblock


\bibitem[Zheng et~al\mbox{.}(2018)]%
        {zheng2018drn}
\bibfield{author}{\bibinfo{person}{Guanjie Zheng}, \bibinfo{person}{Fuzheng Zhang}, \bibinfo{person}{Zihan Zheng}, \bibinfo{person}{Yang Xiang}, \bibinfo{person}{Nicholas~Jing Yuan}, \bibinfo{person}{Xing Xie}, {and} \bibinfo{person}{Zhenhui Li}.} \bibinfo{year}{2018}\natexlab{}.
\newblock \showarticletitle{DRN: A deep reinforcement learning framework for news recommendation}. In \bibinfo{booktitle}{\emph{Proceedings of the 2018 World Wide Web Conference}}. \bibinfo{pages}{167--176}.
\newblock


\bibitem[Zhou et~al\mbox{.}(2023)]%
        {zhou2023adaptive}
\bibfield{author}{\bibinfo{person}{Huachi Zhou}, \bibinfo{person}{Hao Chen}, \bibinfo{person}{Junnan Dong}, \bibinfo{person}{Daochen Zha}, \bibinfo{person}{Chuang Zhou}, {and} \bibinfo{person}{Xiao Huang}.} \bibinfo{year}{2023}\natexlab{}.
\newblock \showarticletitle{Adaptive popularity debiasing aggregator for graph collaborative filtering}. In \bibinfo{booktitle}{\emph{Proceedings of the 46th International ACM SIGIR Conference on Research and Development in Information Retrieval}}. \bibinfo{pages}{7--17}.
\newblock


\bibitem[Zhu et~al\mbox{.}(2021a)]%
        {zhu2021dynamic}
\bibfield{author}{\bibinfo{person}{Ziwei Zhu}, \bibinfo{person}{Yun He}, \bibinfo{person}{Xing Zhao}, {and} \bibinfo{person}{James Caverlee}.} \bibinfo{year}{2021}\natexlab{a}.
\newblock \showarticletitle{Popularity bias in dynamic recommendation}. In \bibinfo{booktitle}{\emph{Proceedings of the 27th ACM SIGKDD Conference on Knowledge Discovery \& Data Mining}}. \bibinfo{pages}{2439--2449}.
\newblock


\bibitem[Zhu et~al\mbox{.}(2021b)]%
        {zhu2021popularity}
\bibfield{author}{\bibinfo{person}{Ziwei Zhu}, \bibinfo{person}{Yun He}, \bibinfo{person}{Xing Zhao}, \bibinfo{person}{Yin Zhang}, \bibinfo{person}{Jianling Wang}, {and} \bibinfo{person}{James Caverlee}.} \bibinfo{year}{2021}\natexlab{b}.
\newblock \showarticletitle{Popularity-opportunity bias in collaborative filtering}. In \bibinfo{booktitle}{\emph{Proceedings of the 14th ACM International Conference on Web Search and Data Mining}}. \bibinfo{pages}{85--93}.
\newblock


\bibitem[Zhu et~al\mbox{.}(2021c)]%
        {zhu2021fairness}
\bibfield{author}{\bibinfo{person}{Ziwei Zhu}, \bibinfo{person}{Jingu Kim}, \bibinfo{person}{Trung Nguyen}, \bibinfo{person}{Aish Fenton}, {and} \bibinfo{person}{James Caverlee}.} \bibinfo{year}{2021}\natexlab{c}.
\newblock \showarticletitle{Fairness among new items in cold start recommender systems}. In \bibinfo{booktitle}{\emph{Proceedings of the 44th International ACM SIGIR Conference on Research and Development in Information Retrieval}}. \bibinfo{pages}{767--776}.
\newblock


\end{thebibliography}
\end{document}